\documentclass[11pt]{article}

\usepackage[margin=1in]{geometry}
\usepackage{amsmath,amssymb,amsfonts,amsthm}
\usepackage{mathtools}
\usepackage{bm}
\usepackage{bbm}
\usepackage{hyperref}
\usepackage{graphicx}
\usepackage{enumitem}
\usepackage{booktabs}
\usepackage{microtype}
\usepackage{tcolorbox}
\usepackage{subcaption}
\usepackage{tabularx}
\usepackage{longtable}
\usepackage{caption}

\usepackage{placeins}
\usepackage{flafter}
\usepackage[table]{xcolor}
\definecolor{groupgray}{RGB}{235,235,235} 

\newtcolorbox{block}[1]{
  colback=blue!5!white,
  colframe=blue!75!black,
  title=#1
}
\usepackage{authblk}
\title{Mitigating Adverse Selection in Concentrated Liquidity AMMs with Dynamic Fees: An Agent-Based Model Approach}
\author[1]{Daniele Maria Di Nosse\textsuperscript{\dag}}
\author[1]{Fabrizio Lillo}

\affil[1]{Scuola Normale Superiore, Pisa, Italy}
\date{\today}

\begin{document}

\maketitle
\begin{center}
  \footnotesize
  \dag Corresponding author. Email: \texttt{daniele.dinosse@sns.it}
\end{center}

\begin{abstract}
Automated Market Makers based on concentrated liquidity, such as Uniswap v3, significantly improve capital 
efficiency but expose Liquidity Providers (LPs) to adverse selection costs, formalized as Loss-Versus-Rebalancing 
(LVR). While theoretical literature quantifies these costs, the interplay between realistic blockchain 
microstructure and endogenous pricing mechanisms remains under-explored. This paper develops a granular 
Agent-Based Model of a Uniswap v3 pool interacting with a stochastic reference market governed by 
Heston volatility dynamics. The framework incorporates discrete block propagation, mempool latency, and a 
heterogeneous population of agents, including latency-sensitive arbitrageurs, smart routers, Maximal Extractable Value searchers, and active LPs benchmarked against a frictionless rebalancing strategy. We propose and
evaluate dynamic fee schedules driven by volatility and order-flow toxicity proxies intended to compensate LPs for
adverse-selection losses. Our simulations investigate the conditions under which LPs can achieve positive hedged Profit
and Loss (fees minus LVR). The analysis suggests that dynamic fee adjustments can improve hedged LP profitability mainly
by increasing fee income in states associated with stale-price risk. Depending on the configuration, these rules may also
affect realized LVR, but the current aggregate results support compensation for LVR more directly than a reduction of LVR
itself.
\end{abstract}


\section{Introduction}

Automated market makers (AMMs) have emerged as a key innovation in Decentralized Finance (DeFi), allowing users to trade assets 
through liquidity pools instead of traditional order books. From a market-microstructure perspective, AMMs can be interpreted
as automated dealer markets in which the fee plays a role analogous to a spread or transaction cost, while inventory adjustment
is embedded in the pricing rule rather than in discretionary dealer quotes \cite{ho1983,avellaneda2008,angeris2021analysis,CARTEA2025105134}.
In AMMs like Uniswap, traders swap against pooled assets and pay fees to liquidity providers (LPs) who supply those assets.
Uniswap v3 introduced concentrated liquidity, enabling LPs
to allocate capital within specified price ranges to improve fee earnings and capital efficiency \cite{adams2021}. This 
advancement comes with increased exposure to market risk: if the price moves outside an LP’s range, the position can 
become effectively inactive and fully converted to one asset. Thus, while concentrated liquidity boosts potential fee 
revenue, it also exacerbates the impermanent loss risk that can be expressed as the opportunity cost or loss an LP 
incurs compared to simply holding the assets. A central question is whether the fees earned by providing liquidity 
are sufficient to compensate for these risks, i.e. whether liquidity provision is profitable net of hedging costs.

Recent research \cite{milionis2023} has formalized the notion of LP profitability by introducing the concept of “hedged 
PnL”, which isolates the profit-and-loss of an LP from the market risk. In particular, the authors in\cite{milionis2023} define the 
delta-hedged LP return as the difference between the so-called \textit{loss-versus-rebalancing} (LVR) cost and the trading 
fees earned \cite{milionis2023}. LVR is the loss an LP suffers due to adverse selection: because AMMs execute trades at 
stale prices, informed arbitrageurs extract value when prices move, leaving LPs worse off than if they had continuously 
rebalanced their portfolio at market prices. This adverse-selection interpretation parallels the classical market-making
literature, where liquidity suppliers require compensation for inventory risk and informed order flow through spreads and
execution premia \cite{ho1983,avellaneda2008,cartea2015book,gueant2016book}. In effect, LVR measures the gap between the LP's value and the value of a dynamically rebalanced benchmark strategy. This benchmark is distinct from the HODL benchmark used to define impermanent loss. The hedged PnL, given by fees earned minus LVR, measures liquidity-provision economics relative to this rebalancing benchmark after separating directional price exposure; it does not imply that stale-quote adverse-selection losses disappear. We therefore keep three benchmarks conceptually separate: unhedged LP returns, HODL-relative performance or impermanent loss, and rebalancing-relative hedged PnL.

The profitability of liquidity provision in concentrated AMMs is not merely a theoretical concern. It has significant 
practical implications for traders, investors, and protocol designers. Empirical evidence has raised concerns that, for 
many volatile asset pools, LPs on Uniswap v3 underperform a buy and holding strategy. Depending on the benchmark considered, LP underperformance may be characterized through impermanent loss relative to HODL or through LVR relative to a rebalancing strategy; these are different comparisons and should not be added without an explicit decomposition. For example, an in-depth analysis of Uniswap v3 by Heimbach et 
al. \cite{heimbach2022} found that across major volatile pools (like ETH-USD or BTC-ETH pairs), LPs on average earned 
negative returns relative to holding, especially when they provided very narrow (concentrated) ranges. The mean daily 
returns of LP positions in such pools were below zero, indicating that fees did not fully offset the losses from adverse 
price movements and arbitrage. Only in low-volatility stablecoin-stablecoin pools did LPs consistently earn small 
positive returns, as impermanent loss was negligible in those cases. Likewise, a comprehensive empirical study by 
Fritsch and Canidio \cite{fritsch2024} examined many large pools and concluded that losses to arbitrageurs (LVR) 
exceeded fee revenues in most cases, resulting in negative hedged PnL for LPs on those pools. The empirical evidence is
therefore consistent with the theoretical view that AMM fees must compensate not only for price volatility, but also for the
timing and informational content of order flow \cite{milionis2023,fritsch2024,maire2024}. Notably, they report
that Uniswap v3’s concentrated liquidity pools were less profitable for passive LPs than the earlier Uniswap v2 
pools, despite the higher fee potential, due to the greater arbitrage losses in v3. These findings underscore that 
liquidity provision, particularly in volatile markets, can be an adverse trade for uninformed or passive providers. 
It also highlights the importance of active management or hedging strategies to improve outcomes.

Against this backdrop, understanding the drivers of LP profitability and exploring methods to enhance it has become a 
multidisciplinary endeavor. On one hand, analytical models from finance can quantify the expected costs and returns of 
liquidity provision under various conditions. On the other hand, Agent-Based Models (ABMs) and simulations offer a way 
to study complex interactions in AMM ecosystems between various kinds of agents that are difficult to capture with 
closed-form theory. The benefit of ABMs is their description of complex systems from the bottom up by specifying the 
behaviour (often very simple) of individual agents and letting aggregate properties emerge from their interactions. 
The main drawback, however, lies in the large number of parameters on which these models typically rely, whose 
calibration is often nontrivial. Broadly speaking, ABM can be interpreted in two distinct ways. They can be viewed 
either as data-generating mechanisms designed to reproduce as many stylized facts as possible (for example, as 
synthetic market generators useful for testing trading strategies) or they can serve as controlled and realistic 
laboratories in which the researcher studies the qualitative response of the system to variations in specific 
parameters (for instance, how a change in tick size affects the price dynamics of a financial asset). This use of ABMs
as controlled laboratories is well established in artificial financial-market research, where heterogeneous-agent simulations
have been used to study transaction taxes, market microstructure, high-frequency trading, cancellation rules, fragmentation,
and the propagation of external information \cite{raberto2001,chakraborti2011,westerhoff2006,mannaro2008,pellizzari2009,leal2016,carro2015,mizuta2020}.
In this paper, we adopt the latter perspective. Our goal is to investigate whether, and in what way, dynamically adjusted fee schedules
influence the economics of liquidity provision in a Uniswap v3–like AMMs. In this sense, AMM fee adaptation is the DeFi analogue
of a market-design intervention: changing a transaction cost or spread-like rule and observing how liquidity, routing, arbitrage,
and welfare-relevant outcomes respond in a heterogeneous-agent system \cite{westerhoff2006,mannaro2008,pellizzari2009,mizuta2020}.

This paper presents an ABM for a Uniswap v3-style pool that allows agents to interact each other, making decisions based 
on the state of the pool and a reference market, a Centralized Exchange (CEX). Specifically, the model incorporates:
\begin{itemize}[leftmargin=*]
  \item tick-based concentrated liquidity;
  \item a reference CEX price evolving according to a Heston volatility dynamics with permanent price impact due 
  to arbitrageurs;
  \item a mempool-based execution;
  \item heterogeneous agents: arbitrageurs, noise traders, liquidity providers and Maximal Extractable Value (MEV) searchers;
  \item passive and active LPs with budgets, review clocks, and risk management;
  \item dynamic taker fees driven by volatility or order flow toxicity;
  \item a rebalancing benchmark for each LP that enforces wealth conservation and allow for a clean hedged-vs-unhedged 
  PnL decomposition;
\end{itemize}

The primary objective is to understand how (and if) dynamical liquidity taker fees can be beneficial for liquidity provision, 
reducing LVR and/or increasing fee revenue for the LPs.

These findings extend the analytical and empirical AMM literature on predictable loss, LVR, and arbitrage losses by studying
how adaptive fee rules interact with block-time execution and strategic liquidity provision
\cite{milionis2023,CARTEA2025105134,fritsch2024}. The simulations deliver three main findings. First, under block-time execution, static fees are insufficient to compensate
standing LPs for stale-quote adverse selection, especially when liquidity is concentrated. Second, dynamic fees can improve
hedged LP profitability, but the aggregate decomposition shows that this improvement is primarily due to higher fee income
rather than necessarily lower cumulative LVR. Rules that react to the contemporaneous gap between the AMM price and the
external reference price perform better in our experiments than rules based only on recent price volatility. Third, fee
adaptivity creates an endogenous trade-off: it compensates standing liquidity by widening the fee wedge, but it also
reduces smart-router DEX share and, when very short-lived liquidity provision is allowed, can redirect part of the enlarged
fee pool toward latency-sensitive liquidity providers.

The paper is organized as follows. Section \ref{sec:literature_review} provides a brief literature review on liquidity provision in AMMs, with particular attention to LP profitability, adverse selection, and dynamic fee mechanisms. Section \ref{sec:model} presents the agent-based model, describing the Uniswap v3 market mechanism, the reference market, the LP PnL accounting framework, the agent population, and the dynamic fee schedules considered in the analysis. Section \ref{sec:results} reports the main simulation results, including the microstructure diagnostics and the comparison across fee policies and model specifications. Finally, Section 5 concludes and discusses possible directions for future research.

\section{Literature Review}
\label{sec:literature_review}
This section reviews the main strands of literature relevant to our research question. We first discuss analytical contributions that formalize LP profitability and adverse selection in AMMs, then summarize the empirical evidence on realized LP performance, and then examine simulation-based and agent-based approaches for AMMs. We then add a broader perspective on ABMs as tools for market-design and policy analysis \cite{raberto2001,chakraborti2011,mizuta2020}. We conclude with a brief overview of academic and deployed dynamic fee mechanisms, since these designs provide useful guidance for the fee rules analyzed in our model.

\subsection{Analytical and Theoretical Approaches}
The foundational models of AMMs and liquidity provider returns have benefited from insights in financial economics. A seminal theoretical contribution is the Black-Scholes-style model for AMM liquidity proposed by Milionis et al. (2023), which explicitly quantifies LP returns and the LVR cost in continuous time \cite{milionis2023}. By modeling the AMM as replicating a continuously rebalanced portfolio, they derive closed-form expressions for LVR under diffusion assumptions. Their model shows that an LP's position can be viewed analogously to writing options: the LP continuously ``sells low and buys high'' when prices move, akin to being short volatility. This explains why, even after hedging out directional price risk, LPs face a systematic shortfall (LVR). The analytical result $\text{Hedged PnL} = 	\text{Fees} - \text{LVR}$ summarizes how fees must overcome LVR for an LP to profit \cite{milionis2023}.

This framing connects AMM liquidity provision to the broader market-making literature, where spreads and execution policies compensate dealers for inventory risk, asymmetric information, and adverse selection \cite{ho1983,avellaneda2008,cartea2015book,gueant2016book}. Cartea, Drissi, and Monga study AMM execution, speculation, predictable loss, and optimal liquidity provision, providing a direct bridge between DeFi mechanisms and continuous-time market-making models \cite{CARTEA2025105134}. Their work ties into the idea of setting optimal trading fees to balance trade frequency and arbitrage loss: higher fees protect LPs from being picked off by arbitrage (reducing LVR), but too high fees reduce volume and fee income. More generally, equilibrium and stochastic-control formulations reinforce the same trade-off: without an appropriate fee or execution design, liquidity-provider returns tend to be squeezed by informed trading and arbitrage in volatile markets.

A distinct line of theoretical inquiry addresses strategies for mitigating impermanent loss. Some works explore hedging impermanent loss using financial derivatives or alternative mechanisms. For example, Maire and Wunsch \cite{maire2024} propose a market-neutral liquidity provision strategy where the LP simultaneously holds an appropriate portfolio of futures/options to offset AMM exposure. While this approach can eliminate directional risk and LVR, they find it requires substantial extra capital and its efficacy depends on market conditions. Another idea is to redesign the AMM mechanism itself; Canidio and Fritsch examine batched trading/auctions as a protocol design to reduce LVR and sandwich attack risks, thereby improving LP outcomes \cite{canidio2025}. These analytical studies contribute to a broader understanding that LP profitability is fundamentally tied to market microstructure – the timing of price updates, arbitrage speed, fee levels, and volatility all determine whether fees outweigh the inherent losses from providing liquidity.

\subsection{Empirical Studies}
Empirical research has leveraged blockchain data to measure realized LP performance in AMMs. Using 
transaction and pool data from Uniswap and other platforms, researchers can compute the actual returns 
of LP positions over time and compare them to benchmarks. The aforementioned study by Heimbach et 
al. \cite{heimbach2022} is one of the first comprehensive looks at Uniswap v3 LP data. They tracked 
individual LP positions in various pools and calculated their holding-period returns, inclusive of 
fees and impermanent loss, versus just holding assets. Their findings were striking: for volatile pairs 
like WETH/USDC or WBTC/WETH, a majority of LPs would have been better off not providing liquidity. 
Narrow-range positions (which concentrate liquidity aggressively) exhibited higher fee income but 
also much higher impermanent losses, often leading to negative net returns on average. By contrast, 
in stablecoin pools (e.g. DAI/USDC), volatility is so low that impermanent loss was minimal and nearly 
all LPs earned small positive returns (albeit with lower fee yields). Heimbach et al. also observed 
that LP positions with very short lifetimes saw the most extreme outcomes – some made high returns by 
timing volatility/volume spikes correctly, while many others lost significantly – suggesting that 
active management and informed timing are crucial for success in Uniswap v3’s environment.

Subsequent empirical studies have reinforced these conclusions. Fritsch and Canidio empirically 
quantify the total dollar value that flowed from LPs to arbitrageurs due to LVR across many Uniswap 
pools \cite{fritsch2024}. They find that this arbitrage extraction often exceeded the fees paid to 
LPs, resulting in net losses. Their work also highlights chain speed as a factor: on faster blockchains 
(or hypothetical faster Ethereum), the arbitrage gaps – and hence LVR – would shrink, improving LP 
outcomes (since prices update closer to continuously). Overall, empirical evidence paints a cautious picture: 
passive liquidity provision is often a negative expected value proposition for volatile asset pairs, whereas 
for more stable pairs or with very active management (frequently rebalancing ranges), LPs can achieve modest 
gains.

\subsection{Agent-Based Modeling and Simulations}
To capture the complex dynamics of AMM ecosystems, researchers have increasingly turned to agent-based models 
(ABM) and simulations. In an ABM, one can simulate a population of agents with different roles – e.g. 
market-makers (LPs), traders, arbitrageurs – interacting according to set rules. This approach, common 
in econophysics, allows experimentation with various scenarios (different volatility regimes, fee levels, 
liquidity distributions, agent strategies, etc.) and observation of emergent outcomes like LP profitability 
or market stability.

One early example of applying simulations to AMMs is by Angeris et al. \cite{angeris2021analysis}, who not only 
provided a theoretical analysis of Uniswap but also created an agent-based simulation to test market 
conditions and arbitrage behavior in a Uniswap pool. Their simulation (using the Gauntlet platform) included 
agents trading and arbitrageurs ensuring prices don’t deviate from an external reference. The results 
supported theoretical properties (e.g. Uniswap’s pricing converges to market price via arbitrage) and 
hinted at how LP returns depend on volatility and arbitrage frequency, laying groundwork for later studies.

A more focused ABM study on LP profitability is by Cohen et al. \cite{cohen2023}. They first derive a 
theoretical upper bound on fees that would be required for an LP to fully hedge away market risk in a 
constant-function market maker (CFMM) like Uniswap. They show analytically that, under typical conditions, 
the required fee to break even (ignoring risk premia) is quite high, hinting that standard fee levels 
(e.g. 0.3\%) might be insufficient. They then implement a multi-agent simulation with LPs and traders, 
varying parameters such as the fee level, asset volatility, and the rate of trade (liquidity taker arrival). 
The simulation confirms that on average, fee income is insufficient to cover the hedging cost for LPs, 
especially as volatility increases or trade frequency decreases \cite{cohen2023}. In their experiments, 
even raising fees does not fully fix the problem: while higher fees reduce LVR (by dissuading marginal 
arbitrage trades), they also reduce volume. The ABM provides a nuanced view of how different factors trade 
off, illustrating, for example, that at very high volatility, no reasonable fee can protect LPs, whereas 
at low volatility, even a small fee can yield positive expected hedged PnL. This aligns with the empirical 
notion that LP profitability is highly regime-dependent.

Another agent-based approach is seen in Hafner and Dietl \cite{hafner2024}. They conduct simulations to 
determine under what conditions an LP’s position outperforms a holding strategy. Their model iterates over 
one-year periods with agents simulating trades consistent with historical Uniswap volumes and volatility. 
Intriguingly, their findings suggest that if the asset’s price fluctuations are not too extreme (for example, 
not more than a -75\% drop or +300\% increase over the year) and the fee volume is substantial, then an 
LP could realize a net profit relative to simply holding the assets \cite{hafner2024}. Essentially, in moderate 
market conditions with decent trading activity, fee accumulation can outpace impermanent loss. However, when 
markets are highly volatile (large price divergence) without commensurate volume, the LP is likely to 
underperform holding. This result offers a more optimistic counterpoint, indicating that there are realistic 
scenarios (especially for less volatile assets or shorter time horizons) where liquidity provision pays off. 
It underscores the importance of market environment in any evaluation of LP profitability.

Agent-based models have also been used to design and test strategic LP behaviors. Rather than treating LPs 
as passive liquidity suppliers, some works model them as active agents who can adjust their liquidity ranges 
withdraw/react based on market signals. Fan et al. \cite{fan2023} present a game-theoretic and computational 
approach for dynamic liquidity provision in Uniswap v3. They consider a setting with one strategic LP who 
periodically updates their price range to maximize earnings, anticipating price moves. By simulating price paths and using techniques like reinforcement learning or neural-network-based optimization, they show that active strategies (like regularly resetting liquidity ranges based on volatility or price trends) can significantly improve an LP’s PnL compared to a static strategy. Their work, along with others, suggests that liquidity provision in Uniswap v3 is an active trading strategy akin to market-making in traditional markets – requiring optimization and sometimes rapid response to market changes. This blurs the line between LPs and traders: the most successful LPs behave like informed agents, adjusting positions to manage inventory and risks.

Incorporating hedging into agent strategies is another frontier. Zhang et al. \cite{zhang2023} employ deep 
reinforcement learning (DRL) to train an agent that provides liquidity in Uniswap v3 while simultaneously 
hedging with external instruments (like futures) to offset price risk . The DRL agent learns to shift its 
liquidity range and maintain an appropriate hedge ratio in a way that maximizes the expected hedged PnL 
(trading off fees earned against hedging costs and gas fees). In simulations on volatile pairs like ETH/USDC, 
the learned strategy outperformed naive strategies and achieved better risk-adjusted returns \cite{zhang2023}. 
This approach illustrates the potential of ABM combined with machine learning: one can simulate a realistic 
market environment (with price paths, trading costs, etc.), train agent strategies within it, and derive 
insights on best practices for LPs. Notably, Zhang et al. confirm that hedging significantly reduces PnL 
volatility (by removing most impermanent loss), and the remaining key to profitability is optimizing fee 
collection versus transaction costs. Their agent effectively implements the delta-hedged LP strategy in an 
automated fashion, reinforcing the concept that hedged PnL is the right objective for liquidity providers.

In summary, the literature on liquidity provision profitability spans analytical finance, empirical data 
analysis, and agent-based simulation. The consensus from these varied approaches is that liquidity provision 
in concentrated AMMs is a high-risk, low-margin endeavor for most typical users, especially in volatile 
markets. Adverse selection (LVR) tends to systematically eat into LP returns, meaning that without either 
specialized strategies or favorable market conditions, LPs may not be adequately compensated for the risk. 
Agent-based models have been particularly useful in stressing the system under different conditions and 
testing potential improvements. They have shown, for instance, that protocol design changes (like fee 
adjustments or batch auctions) can alter the balance of profits, and that sophisticated, active strategies 
(possibly aided by algorithmic agents or automation) are necessary to achieve positive expected returns. 
From a finance perspective, this aligns with the idea that passive market-making can be a losing proposition 
when trading with informed counterparts – an insight that echoes traditional market microstructure theory 
(where market makers require a spread or other edge to profit). In the context of AMMs, the “spread” is 
essentially the fee, and current fee levels often appear too low to offset the adverse selection cost under 
fast-moving prices.

\subsection{ABMs for Market Design and Financial-Policy Analysis}
A related strand of artificial financial-market research uses ABMs to evaluate market-design interventions. 
Early artificial markets showed how stylized price dynamics can emerge from heterogeneous 
interacting agents \cite{raberto2001,chakraborti2011}. Subsequent work uses ABMs to study 
how transaction taxes and fee-like frictions affect volatility, liquidity, and market stability under 
different microstructures \cite{westerhoff2006,mannaro2008,pellizzari2009}.Other studies 
examine how rule changes related to hedging instruments, high-frequency trading, order cancellation, latency, 
and external information can amplify or dampen market instability \cite{brock2009,leal2016,carro2015,gao2024,
yagi2020}. Our use of an ABM to test dynamic AMM fees follows this market-design logic: the 
fee rule is treated as a policy lever whose impact depends on heterogeneous agent responses and feedback 
through prices, routing, liquidity provision, and arbitrage \cite{mizuta2020}.

\subsection{Dynamic fees in deployed AMMs}
\label{subsec:deployed_dynamic_fees}

Academic work on AMM fee choice emphasizes that the fee is not merely a revenue parameter but a control variable that trades off volume against adverse selection and arbitrage losses \cite{milionis2023,CARTEA2025105134}. Deployed protocols implement this idea through simpler, bounded controllers based on imbalance, volatility, or activity proxies.

Beyond academic proposals, several large AMM families already deploy fee schedules that adapt endogenously to risk or imbalance conditions. These implementations are useful reference points because they reveal which signals are considered sufficiently robust to run ``on-chain'' and which functional forms are preferred to keep fees bounded, smooth, and hard to game.

A first design pattern is off-peg / imbalance-based fees, common in stable-asset designs. The basic idea is that, for assets that should trade close to a target relative price, the pool's reserve composition is itself an informative risk signal. When the pool is close to its target balance, trades are likely to be ordinary liquidity-taking flow and can be charged a low baseline fee. When trades push the pool away from balance---for example, leaving it with too much of one asset and too little of the other---the pool is effectively absorbing inventory and de-peg risk. The fee is therefore increased to discourage trades that further worsen the imbalance and to compensate LPs for taking the other side of stressed flow. In Curve's Stableswap-NG, this idea is implemented with a compact balance metric
$K(x,y)=\frac{4xy}{(x+y)^2}\in(0,1]$, where $x$ and $y$ denote normalized reserves and $K=1$ at perfect balance ($x=y$). As the pool becomes more imbalanced, $K$ falls below one and the dynamic fee rises through a rational function that equals the baseline fee at the peg and approaches a capped multiple of it as imbalance grows \cite{curve_stableswap_ng}. Curve v2 (CryptoSwap) implements a closely related idea: the fee interpolates between a ``mid'' fee (near balance) and an ``out-of-range'' fee (far from balance) through a smooth imbalance coefficient, often expressed via a geometric-vs-arithmetic-mean ratio and a reduction factor that shrinks under imbalance \cite{curve_cryptoswap_whitepaper}. For modeling purposes, both mechanisms can be read as: measure how far the pool state is from its target composition with a scalar statistic, then map that statistic monotonically into a higher effective fee, with a hard ceiling.

A second pattern is activity/volatility-accumulator fees, prevalent in ``bin'' AMMs. Trader Joe's Liquidity Book computes fees as the sum of a base term proportional to the bin step and a quadratic variable term driven by a volatility accumulator that grows when swaps traverse bins and decays when activity slows \cite{lfj_fees}. Meteora's DLMM follows the same blueprint (base plus a capped quadratic variable component) with explicit scaling constants and a maximum fee rate \cite{meteora_dlmm_docs}. Conceptually, these designs make fees respond quickly to short-lived bursts of toxicity (rapid price moves and frequent bin crossings), while ensuring mean reversion through time-decay.

A third pattern appears in concentrated-liquidity designs that remain close to the Uniswap v3 paradigm. Algebra's adaptive fee combines (i) a volatility premium computed as a sum of sigmoid responses of a tick-based volatility estimator and (ii) a ``volume-per-liquidity'' regulator, then clamps the result between minimum and maximum fees \cite{algebra_adaptive_fee_docs,algebra_tech_paper}. The practical takeaway is that smooth saturating functions (sigmoids) are used to avoid fee flicker, while multiplicative regulators incorporate local market conditions (how much flow is being processed per unit of liquidity).

Overall, deployed protocols converge on three robust implementation choices: (i) a low-dimensional risk signal (imbalance, volatility proxy, bin-crossing intensity), (ii) a monotone mapping that increases fees in stressed regimes (often convex/quadratic or saturating), and (iii) explicit caps and decay to stabilize dynamics. These motifs directly motivate the simplified controllers we test in Section \ref{subsec:fees}, where fees are updated on a block clock using volatility- and toxicity-proxy signals.

\section{Model}
In this section, we describe the dynamics of the model. We begin by outlining the mechanics of Uniswap v3, followed by a description of how the reference (CEX) market is modeled and how the different cohorts of agents behave. We then introduce three simple dynamic fee schedules: one that adjusts fees in response to CEX or DEX price volatility and one based on a proxy for order-flow toxicity (measured by how much the AMM price is stale during the coupled CEX-DEX dynamics). Finally, we explain how liquidity providers’ PnL is calculated and provide key details of the simulation framework.
\label{sec:model}

\subsection{Uniswap v3 Market Mechanism}
\label{subsec:univ3}

We consider an AMM that allows swapping between two assets, token 0 and token 1, denoted respectively with $x$ and $y$. Throughout, prices are quoted as token 1 per unit of token 0. Uniswap v3 implements a concentrated-liquidity AMM \cite{adams2021}, meaning that liquidity providers (LPs) choose the price range over which their capital is active. Whenever a swap occurs
the liquidity taker (LT) must pay a fee $f_t$ to the LPs. We intentionally put the dependence of the fee on time, since the main objective of this paper is to understand what happens to the economics of the agents if we let the fee have a non trivial dynamics. This section introduces the main objects needed to model Uniswap v3 swaps.

\subsubsection{Discrete price grid, ticks, and square-root price}

Uniswap v3 discretises the price axis into ticks indexed by integers $i \in \mathbb{Z}$.
For simplicity we assume tick spacing $\Delta i = 1$
\footnote{Tick spacing is fixed by the pool contract; e.g., for USDC--WETH at 0.05\% the tick spacing is 10.}.
The price associated with tick $i$ is
\begin{equation}
  p(i) = \gamma^{i}, \qquad \gamma = 1.0001,
\end{equation}
and Uniswap v3 uses the square-root price instead of the simple price because, under constant liquidity, token amounts vary linearly with 
changes in \(s=\sqrt{p}\), which makes swap and position accounting algebraically and computationally simple:
\begin{equation}
  s(i) = \sqrt{p(i)} = \gamma^{i/2}.
\end{equation}
At any time $t$, the pool is at an active tick $i_t$ and active square-root price $s_t$. Between two adjacent tick boundaries, the active liquidity is constant and the swap dynamics admit closed-form updates like in Uniswap v2 \cite{Adams2020UniswapVC}.

\subsubsection{Liquidity positions and active liquidity}

An LP does not provide liquidity over the entire price axis. Instead, an LP position is specified by
a tick interval $[i_a,i_b]$ and a liquidity amount $L^{\mathrm{pos}}>0$.
Define the corresponding boundary square-root prices
\begin{equation}
  s_a := s(i_a), \qquad s_b := s(i_b).
\end{equation}
Intuitively, a position is active only when the current price lies inside its chosen range. When active, it contributes depth to the pool and earns a pro-rata share of swap fees; when inactive, it earns no fees and is held entirely in one of the two tokens.

Let $L_t$ denote the active liquidity at time $t$, i.e., the sum of liquidity across all positions whose ranges contain the active tick:
\begin{equation}
  L_t \;=\; \sum_{\mathrm{pos}} L^{\mathrm{pos}}\,\mathbf{1}\!\left(i_a^{\mathrm{pos}} \le i_t < i_b^{\mathrm{pos}}\right).
\end{equation}
A useful representation of how liquidity changes across the grid is the liquidity net $\ell(i)$, defined as the net change in active liquidity when crossing tick $i$:
\begin{equation}
  \ell(i) \;=\; \sum_{\mathrm{pos}} L^{\mathrm{pos}}\mathbf{1}(i = i_a^{\mathrm{pos}})\;-\;\sum_{\mathrm{pos}} L^{\mathrm{pos}}\mathbf{1}(i = i_b^{\mathrm{pos}}).
\end{equation}
When the price crosses a tick boundary, some positions become active/inactive and the pool updates $L_t$ by adding $\ell(i)$ at that boundary.

\paragraph{Token composition of a position.}
Within its active range ($s_a \le s_t \le s_b$), the amounts of token 0 and token 1 represented by liquidity $L^{\mathrm{pos}}$ at square-root price $s_t$ are
\begin{equation}
  x^{\mathrm{pos}}(s_t) = L^{\mathrm{pos}}\!\left(\frac{1}{s_t} - \frac{1}{s_b}\right),
  \qquad
  y^{\mathrm{pos}}(s_t) = L^{\mathrm{pos}}\!\left(s_t - s_a\right).
\end{equation}
Outside the range, the position is entirely in one asset:
\begin{equation}
  \begin{aligned}
    s_t \le s_a &: \quad x^{\mathrm{pos}} = L^{\mathrm{pos}}\!\left(\frac{1}{s_a} - \frac{1}{s_b}\right), \;\; y^{\mathrm{pos}}=0,\\
    s_t \ge s_b &: \quad x^{\mathrm{pos}} = 0, \;\; y^{\mathrm{pos}}= L^{\mathrm{pos}}\!\left(s_b - s_a\right).
  \end{aligned}
\end{equation}
These relations highlight why liquidity is ``concentrated'': by choosing a narrow interval, an LP increases the effective depth (active liquidity) around that region, at the cost of becoming inactive if the price moves away.

\subsubsection{Swap dynamics within a tick}
Here we briefly describe how a swap works in Uniswap v3.
Consider a swap executed at time $t$ while the price remains within the current tick interval, so that active liquidity $L_t$ is constant. In this regime, Uniswap v3 behaves like a continuous AMM (like Uniswap v2). parameterised by $(s_t, L_t)$, and the input/output amounts are determined by the change in square-root price.

Let the price move from $s$ to $s'$ with constant liquidity $L$ (we omit the time subscript for readability).
The corresponding token amount changes satisfy the standard Uniswap v3 relations:
\begin{equation}
  \Delta x = L\!\left(\frac{1}{s'} - \frac{1}{s}\right),
  \qquad
  \Delta y = L\!\left(s' - s\right).
  \label{eq:univ3_dx_relations}
\end{equation}
The sign and direction of the price move depend on the trade direction:
\begin{itemize}[leftmargin=*]
  \item If a trader swaps token 0 into the pool to receive token 1, the pool's token 0 reserve increases and token 1 reserve decreases, so the price $p=s^2$ moves down (thus $s' < s$). In \eqref{eq:univ3_dx_relations}, this corresponds to $\Delta x>0$ and $\Delta y<0$.
  \item If a trader swaps token 1 into the pool to receive token 0, the price moves up (thus $s' > s$), giving $\Delta y>0$ and $\Delta x<0$.
\end{itemize}

Fees are charged on the input side. If the pool fee rate is $f \in (0,1)$, then for a submitted input amount $\Delta x_{\mathrm{in}}$ only $(1-f)\Delta x_{\mathrm{in}}$ is effectively used to move the price; the remaining $f\Delta x_{\mathrm{in}}$ is accumulated as fees and later distributed to LPs active over the traversed price region, proportional to their share of liquidity.

\subsubsection{Tick crossing and pathwise execution of large swaps}

Because liquidity is concentrated, the pool depth varies across prices. A large swap may move the price across multiple ticks, activating and deactivating different LP ranges. Uniswap v3 therefore executes swaps tick by tick:
within each tick interval the liquidity is constant and \eqref{eq:univ3_dx_relations} applies, while at tick boundaries the liquidity is updated using the liquidity net.

For definiteness, suppose a trader submits an order of size $\Delta x>0$ (token 0 in, token 1 out) at time $t$.
Starting from $(s_t, i_t, L_t)$, the protocol:
\begin{enumerate}[leftmargin=*]
  \item Computes the maximum amount of token 0 that can be consumed before the next tick boundary is reached (given the current $s_t$ and constant $L_t$), accounting for fees on the input.
  \item If the remaining order is below this capacity, it is fully executed within the current tick by updating $s_t$ using \eqref{eq:univ3_dx_relations}, and the swap terminates.
  \item Otherwise, the swap moves the price to the tick boundary, consumes the corresponding capacity in that tick, and crosses into the adjacent tick. At the boundary the active tick index and active liquidity are simultaneously updated as
  \begin{equation}
    i_t \leftarrow i_t + \varepsilon, \qquad
    L_t \leftarrow \begin{cases}L_t +  \ell(i_t+1) \quad &\text{if} \quad  \varepsilon=1 \\
                                L_t -  \ell(i_t) \quad &\text{if} \quad  \varepsilon=-1\end{cases}
  \end{equation}
  where $\varepsilon$ is the sign of the tick move (-1 for token 0 in and +1 for token 1 in).
  \item The procedure repeats until the full order is executed or no active liquidity remains.
\end{enumerate}

This mechanism implies that the swap's price impact is generally piecewise, determined by the sequence of ticks crossed and the corresponding liquidity profile $L$ along the path. Fees are accrued separately within each traversed tick interval and allocated to the LPs whose positions are active over that interval, proportional to their provided liquidity.

\subsubsection{Mempool}
\label{subsubsec:mempool}

The simulation evolves in discrete blocks indexed by $t=1,\dots,T$, mirroring the
block-based execution model of Ethereum-like chains. Each block is subdivided into
$B\ge 1$ micro-steps (treated as seconds) indexed by $\tau=1,\dots,B$. Hence, $B$ represents the block time measured in units of a micro-step. 

At the end of block $t$, the simulator records a validated on-chain snapshot
\[
  S_t \;=\; \bigl(P_t^{\mathrm{DEX}},\, L_t,\, m_t\bigr),
\]
where $P_t^{\mathrm{DEX}}$ is the AMM price, $L_t$ is the active liquidity state, and
$m_t$ is the external CEX mid-price. At the beginning of block $t+1$, all agents
observe $S_t$ and condition their decisions on it.
This captures (i) confirmation latency—agents react to the last confirmed on-chain
state—and (ii) a common reference for evaluating slippage, basis, and range placement
within the block.

During the micro-steps, the external reference price continues to
evolve (in our baseline, via a Heston stochastic-volatility specification; see
Section \ref{subsec:cex}), while the observable on-chain state remains frozen
at $S_{t}$ from the agents' perspective. Agents therefore submit actions using
information that becomes stale by the time it is executed, consistent with
blockchain-style batch settlement.

All intents generated over the $B$ micro-steps are appended to a pending mempool and
are not executed immediately. In parallel, an arbitrage intent is constructed
to exploit the price discrepancy implied by the frozen on-chain snapshot and the
reference  at the beginning of the block. At the block boundary, the simulator executes the block via
a mempool replay:
\begin{enumerate}[leftmargin=*]
  \item Collect all intents accumulated during block $t$.
  \item Randomly shuffle the mempool to represent stochastic transaction ordering (i.e. there is no auction for block space and block position).
  \item Execute the arbitrage intent with priority (to approximate MEV-style
        ordering advantages), then execute the remaining intents sequentially against
        the live AMM state.
  \item Record the resulting state as the next validated snapshot $S_t$.
\end{enumerate}
This mechanism decouples stochastic order arrival from on-chain ordering while keeping
per-block targets interpretable and directly controllable through $\lambda_e$, and it
captures the key latency feature of blockchain systems: agents act on confirmed
information, yet execution occurs later under uncertain ordering, with arbitrageurs
enjoying systematic priority. This design abstracts from the full block-building market, but preserves the market-design feature that latency and priority ordering can materially affect volatility, arbitrage, and liquidity provision, as emphasized in both MEV studies and artificial-market analyses of high-frequency trading and cancellation rules \cite{flashboys,leal2016,gao2024,yagi2020}.

\subsection{Reference Market (CEX)}
\label{subsec:cex}

The reference market is a centralized exchange (CEX) endowed by an underlying price dynamics
and a coupling with the DEX due to the price impact induced by arbitrageurs.
We denote by $m_t$ the mid-price at the end of block $t$. The CEX is used for:
\begin{itemize}[leftmargin=*]
  \item defining the arbitrage band against the DEX;
  \item marking LP wealth and fees to market;
  \item implementing the rebalancing benchmark associated with the Loss-Versus-Rebalancing (LVR).
\end{itemize}

\subsubsection{Heston Volatility Model}

To capture stylized facts of asset returns such as volatility clustering and the leverage effect (the negative correlation 
between returns and volatility changes), we introduce a stochastic volatility regime based on the Heston model \cite{heston1993}. Unlike the
geometric Brownian motion (GBM) with constant or deterministically shifting volatility, the variance $v_t = \sigma_t^2$ in 
this regime follows a mean-reverting stochastic process.

The evolution of the reference market price $m_t$ and its variance $v_t$ is governed by the following system of stochastic 
differential equations (SDEs):

\begin{align}
    d \ln m_t &= \left(\mu - \frac{1}{2}v_t\right)dt + \sqrt{v_t} dW_t^{(1)}, \\
    dv_t &= \kappa(\theta - v_t)dt + \sigma_v \sqrt{v_t} dW_t^{(2)},
    \label{eq:heston_dynamics}
\end{align}

where $W_t^{(1)}$ and $W_t^{(2)}$ are standard Brownian motions with correlation $\rho$, i.e., $\mathbb{E}[dW_t^{(1)} dW_t^{(2)}] = \rho dt$. The parameters are defined as follows:
\begin{itemize}
    \item $\mu$: The drift of the log-price process.
    \item $\theta$: The long-run average variance.
    \item $\kappa$: The rate of mean reversion, determining how quickly $v_t$ returns to $\theta$.
    \item $\sigma_v$: The volatility of volatility (vol-of-vol).
    \item $\rho$: The correlation parameter. A negative $\rho$ induces the leverage effect, where price drops are associated with increases in volatility.
\end{itemize}
The long-run variance has been calibrated using data from Binance regarding the ETH/USD pair, between the 1 January 2023 and 31 December 2023.

\subsubsection{Market Impact}
\label{subsubsec:impact}
Trades performed on the CEX are assumed to have zero costs but permanent price impact on the CEX price.
Let $\Delta a_\tau$ denote the net signed volume in token 0/token 1 executed on the CEX during microstep $\tau$. Empirically, 
large trades tend to move prices in a concave way, consistent with the square-root law observed in market-impact 
studies \cite{PhysRevX.1.021006,Maitrier03042026}, Inspired by these findings, the impact function is specified as
\begin{equation}
  \mathrm{Impact}_\tau =
  \eta_{\text{imp}} \cdot \mathrm{sign}(\Delta a_\tau)\,\sqrt{\bigl|\Delta a_\tau\bigr|},
  \label{eq:impact}
\end{equation}
with $\eta_{\text{imp}}>0$ and $\xi \geq 0$.
From the point of view of the arbitrageur, this ensures that large trades move the CEX against the arbitrageur, 
providing a natural feedback that reduces future mispricing.

\subsection{LVR and PnL Accounting}
\label{sec:lvr}

As we already stated, LVR quantifies the structural cost borne by liquidity providers due to the passive nature of 
Automated Market Makers. It was first introduced by Milionis et al. \cite{milionis2023}. This stale-quote cost is the AMM analogue of adverse selection faced by liquidity suppliers in dealer and limit-order-book markets \cite{ho1983,avellaneda2008,milionis2023,CARTEA2025105134}. Unlike active market
makers on Centralized Exchanges (CEX) who can update quotes in real-time, AMM positions offer stale prices that 
are "picked off" by arbitrageurs whenever the external market moves. LVR measures the cumulative value extracted 
by these arbitrageurs---or equivalently, the opportunity cost of providing liquidity compared to a specific benchmark 
strategy.

To isolate the cost of liquidity provision from market risk, the authors in \cite{milionis2023} compare the LP's wealth 
to a theoretical rebalancing strategy. This benchmark strategy is defined by two key properties:
\begin{enumerate}
    \item Inventory Matching: At every instant $t$, the rebalancing strategy holds the exact same quantity of the risky asset (token 0), denoted by $x_t$, as the actual LP position in the AMM.
    \item Frictionless Execution: Whenever the inventory $x_t$ changes (due to trading activity in the pool), the rebalancing strategy executes the corresponding trade at the fresh, external reference market price $m_t$, rather than at the stale AMM price $P^{DEX}_t$.
\end{enumerate}
Because the rebalancing strategy executes the same sequence of trades as the AMM but at better prices (selling at $m_t > P^{DEX}_t$ when prices rise, and buying at $m_t < P^{DEX}_t$ when prices fall), its value $V^{reb}_t$ strictly exceeds the LP's value $V^{LP}_t$. The difference is the LVR:
\begin{equation}
    LVR_t = V^{reb}_t - V^{LP}_t.
\end{equation}

Milionis et al. \cite{milionis2023} show that if the external reference price $m_t$ follows a diffusion process with 
instantaneous volatility $\sigma_t$, LVR accumulates as a function of the price volatility and the AMM's marginal liquidity. 
The cumulative LVR over a period $[0, T]$ is given by:
\begin{equation}
    LVR_{[0, T]} = \int_0^T \frac{\sigma_t^2 m_t^2}{2} |x^{*\prime}(m_t)| \, dt,
    \label{eq:lvr_general}
\end{equation}
where $x^*(p)$ is the AMM's demand curve (the inventory of token 0 held by the pool at price $p$), and $|x^{*\prime}(p)| = |\frac{dx}{dp}|$ is the marginal liquidity, representing how aggressively the AMM trades against price changes.

In the specific context of Uniswap v3 , the inventory function is determined by the active concentrated liquidity $L_t$. 
Within a specific tick range, the relationship between token 0 inventory and square-root price $s = \sqrt{p}$ is 
given by Equation (7):
\begin{equation}
    x(s) = L_t \left( \frac{1}{s} - \frac{1}{s_b} \right).
\end{equation}
To apply the general LVR Equation \ref{eq:lvr_general}, we compute the marginal liquidity with respect to the price $p = s^2$. Using the chain rule:
\begin{equation}
    \left| \frac{dx}{dp} \right| = \left| \frac{dx}{ds} \frac{ds}{dp} \right| = \left| \left( -\frac{L_t}{s^2} \right) \left( \frac{1}{2\sqrt{p}} \right) \right| = \frac{L_t}{2 p \sqrt{p}}.
\end{equation}
Substituting this marginal liquidity back into Equation \ref{eq:lvr_general} (with $p = m_t$), we obtain the instantaneous LVR specific to a Uniswap v3 position:
\begin{equation}
    l_t^{v3} = \frac{\sigma_t^2 m_t^2}{2} \cdot \frac{L_t}{2 m_t \sqrt{m_t}} = \frac{\sigma_t^2 L_t \sqrt{m_t}}{4}.
    \label{eq:lvr_univ3}
\end{equation}
Equation \ref{eq:lvr_univ3} provides the key intuition for our agent-based model: the cost of providing liquidity scales linearly with the active liquidity $L_t$ and the price level $\sqrt{m_t}$, and quadratically with volatility $\sigma_t$. Crucially, because $L_t$ is the concentrated liquidity, LVR is only incurred when the reference price $m_t$ falls within the LP's active tick range (where $L_t > 0$); inactive positions do not suffer LVR.

\subsubsection{Rebalancing Benchmark in Discrete Time}

The ABM computes LVR using a discrete-time rebalancing benchmark that faithfully
tracks the LP's token exposure.

Fix an LP and let $m_t$ be the CEX mid at the end of block $t$.
Let $x_t$ denote the LP's total exposure in token 0 at time $t$ (the sum over all its AMM
positions) and $y_t$ its cash holdings in token 1. The LP's mark-to-market wealth is
\begin{equation}
  V_t^{\mathrm{LP}} = x_t m_t + y_t.
\end{equation}

We construct a benchmark portfolio $(x_t^{\mathrm{reb}}, y_t^{\mathrm{reb}})$ with the
following properties:
\begin{itemize}[leftmargin=*]
  \item it always holds the same exposure in token 0 as the LP,
        \begin{equation}
          x_t^{\mathrm{reb}} = x_t;
        \end{equation}
  \item it trades only at the CEX mid $m_t$;
  \item it is self-financing.
\end{itemize}

Let the benchmark be initialized at $t=0$ with
\begin{equation}
  x^{\mathrm{reb}} = x,\qquad
  y_0^{\mathrm{reb}} = y_0,\qquad
  V_0^{\mathrm{reb}} = V_0^{\mathrm{LP}}.
\end{equation}

Then we update it in two ways:

\paragraph{Price moves (passive holding).}
When the CEX price moves from $m_{t-1}$ to $m_t$ while the benchmark holds
$x_{t-1}^{\mathrm{reb}}$ units of token 0, the benchmark's value changes by
\begin{equation}
  \Delta V_t^{\mathrm{reb},\mathrm{price}} =
  x_{t-1}^{\mathrm{reb}} (m_t - m_{t-1}),
\end{equation}
so that
\begin{equation}
  V_t^{\mathrm{reb}} = V_{t-1}^{\mathrm{reb}}
                       + x_{t-1}^{\mathrm{reb}} (m_t - m_{t-1}).
\end{equation}

\paragraph{Rebalancing trades (active adjustments).}
Whenever the LP's exposure changes from $x_{t-}^{\mathrm{LP}}$ to $x_{t+}^{\mathrm{LP}}$
due to AMM interactions (minting, burning, price moves within the pool), the benchmark
trades at the CEX price $m_t$ to match the new exposure:
\begin{equation}
  x_{t+}^{\mathrm{reb}} = x_{t+}^{\mathrm{LP}},\qquad
  y_{t+}^{\mathrm{reb}} = y_{t-}^{\mathrm{reb}}
                         - \bigl(x_{t+}^{\mathrm{reb}} - x_{t-}^{\mathrm{reb}}\bigr) m_t.
\end{equation}
This ensures the benchmark is self-financing:
changes in $x^{\mathrm{reb}}$ are funded by opposite changes in $y^{\mathrm{reb}}$, with no
external cash injections.

By construction, the benchmark value is
\begin{equation}
  V_t^{\mathrm{reb}} = x_t^{\mathrm{reb}} m_t + y_t^{\mathrm{reb}}.
\end{equation}

\subsection{Agents}
\label{sec:agents}

The ecosystem comprises several types of agents with distinct objectives, information sets, and constraints. In this section, we describe their nature and behaviour.

\subsubsection{Arbitrageurs}

Arbitrageurs try to keep the price discrepancy between the DEX and the CEX price as small as possible, up to wedges induced by the AMM taker fee and flash-loan funding.
At the beginning of each block they observe the validated snapshot $(m_t, P_t^{\mathrm{DEX}})$,
where $m_t$ is the CEX mid-price (token 1 per token 0) and $P_t^{\mathrm{DEX}}$ is the current
marginal DEX price.

Let $f_t\in[0,1)$ denote the taker fee at time $t$ and $\phi_{\mathrm{flash}}\ge 0$ the flash-loan fee,
interpreted as a proportional cost on the borrowed principal. Arbitrage is modeled as:
(i) borrow the input token via flash loan; (ii) swap against the AMM; (iii) unwind on the CEX at price
$m_t$ to repay principal plus fee. All payoffs are measured in token 1 units.
Because the AMM price moves during execution, profitability is inherently size-dependent.
The arbitrageur therefore chooses an executed swap size $q\ge 0$ to maximize net profit. The arbitrageur is therefore both the mechanism enforcing price alignment and the channel through which stale AMM quotes become realized LP losses \cite{angeris2021analysis,milionis2023,fritsch2024,CARTEA2025105134}. We have two possible situations, described below.

\paragraph{DEX overpriced: token 0 in.}
If $P_t^{\mathrm{DEX}}$ is high relative to $m_t$, the arbitrageur borrows token 0, inputs
$q$ token 0 into the AMM, receives an output $Y_t(q)$ in token 1, and repays $(1+\phi_{\mathrm{flash}})q$
token 0 by buying it on the CEX at price $m_t$.
Net profit is
\begin{equation}
  \Pi_t^{(0\rightarrow 1)}(q)
  \;=\;
  Y_t(q) \;-\; m_t(1+\phi_{\mathrm{flash}})\,q \;-\; c_{\mathrm{fix}},
  \label{eq:arb-profit-0in}
\end{equation}
where $c_{\mathrm{fix}}\ge 0$ captures fixed costs (e.g. gas) in token 1 units. For simplicity, we will
consider $c_{\mathrm{fix}} = 0$. 

Fees are charged on the input: only the fee-adjusted amount $u=(1-f_t)q$
moves the AMM state. Let $P_t(q)$ denote the marginal DEX price (token 1 per token 0)
after executing size $q$ (piecewise across ticks). For an infinitesimal increase $dq$, the AMM output
satisfies
\begin{equation}
  \frac{dY_t(q)}{dq} = (1-f_t)P_t(q).
  \label{eq:dYdq}
\end{equation}
Hence the marginal net profit is
\begin{equation}
  \frac{d\Pi_t^{(0\rightarrow 1)}(q)}{dq}
  =
  (1-f_t)\,P_t(q)- m_t(1+\phi_{\mathrm{flash}}),
  \label{eq:marginal-0in}
\end{equation}
and an interior optimum $q^*>0$ satisfies the first-order condition
\begin{equation}
  P_t(q^*) = \frac{m_t(1+\phi_{\mathrm{flash}})}{1-f_t}.
  \label{eq:target-price-upper}
\end{equation}
Operationally, this means the arbitrageur trades token 0 into the AMM until the marginal DEX price is
pushed down to the boundary \eqref{eq:target-price-upper}, or until liquidity is exhausted.

\paragraph{DEX underpriced: token 1 in.}
If $P_t^{\mathrm{DEX}}$ is low relative to $m_t$, the arbitrageur borrows token 1, inputs $q$ token 1
into the AMM, receives an output $X_t(q)$ in token 0, sells it on the CEX for $m_t X_t(q)$ token 1, and
repays $(1+\phi_{\mathrm{flash}})q$ token 1. Net profit is
\begin{equation}
  \Pi_t^{(1\rightarrow 0)}(q)
  =
  m_tX_t(q) - (1+\phi_{\mathrm{flash}})q - c_{\mathrm{fix}}.
  \label{eq:arb-profit-1in}
\end{equation}
Within the AMM, the fee-adjusted input is again $u=(1-f_t)q$. Let $P_t(q)$ denote the marginal DEX price
after executing size $q$ (now increasing in $q$). The marginal token 0 output per unit token 1 input is
the reciprocal price, so
\begin{equation}
  \frac{dX_t(q)}{dq}
  =
  (1-f_t)\frac{1}{P_t(q)}.
  \label{eq:dXdq}
\end{equation}
Thus
\begin{equation}
  \frac{d\Pi_t^{(1\rightarrow 0)}(q)}{dq}
  =
  m_t(1-f_t)\frac{1}{P_t(q)} - (1+\phi_{\mathrm{flash}}),
  \label{eq:marginal-1in}
\end{equation}
and an interior optimum $q^*>0$ satisfies
\begin{equation}
  P_t(q^*) = \frac{m_t(1-f_t)}{1+\phi_{\mathrm{flash}}}.
  \label{eq:target-price-lower}
\end{equation}
That is, the arbitrageur trades token 1 into the AMM until the marginal DEX price is pushed up to the
boundary \eqref{eq:target-price-lower}, or until liquidity is exhausted.

\paragraph{No-arbitrage band and execution rule}

Equations \eqref{eq:target-price-lower}--\eqref{eq:target-price-upper} imply a two-sided no-arbitrage band
(in marginal-price terms):
\begin{equation}
  \frac{m_t(1-f_t)}{1+\phi_{\mathrm{flash}}}
  \le
  P_t^{\mathrm{DEX}}
  \le
  \frac{m_t(1+\phi_{\mathrm{flash}})}{1-f_t}.
  \label{eq:no-arb-band}
\end{equation}
If $P_t^{\mathrm{DEX}}$ lies inside \eqref{eq:no-arb-band}, the arbitrageur does not trade.
Otherwise, it computes the executed size $q^*$ by simulating the AMM swap (accounting for the input fee)
until the marginal DEX price reaches the nearest boundary
\eqref{eq:target-price-lower} or \eqref{eq:target-price-upper}, or until no further progress is possible
due to finite liquidity. The trade is executed only if the corresponding net profit
$\Pi_t^{(0\rightarrow 1)}(q^*)$ or $\Pi_t^{(1\rightarrow 0)}(q^*)$ is strictly positive, in which case
flash-loan costs (and fixed costs) generate a non-zero region of persistent mispricings. The band in \ref{eq:no-arb-band} is similar to the one derived in \cite{angeris2021analysis} and it becomes exactly equal if $\phi_{\text{flash}}=0$. Our setting, however, improve the realism of these band by introducing frictions via the flash-loan fees and finite liquidity.

\paragraph{Execution and Impact}

The arbitrageur trades against the pool
to move the DEX price as close as possible to the relevant band boundary in \eqref{eq:no-arb-band},
subject to liquidity depth. It may not always be possible to reach the band exactly; in that case,
the arbitrageur consumes all available depth in the profitable direction and stops.

Arbitrage trades are assumed to incur CEX impact as in \eqref{eq:impact}. In the baseline regime,
apart from this permanent impact and the flash-loan fee, arbitrageurs are:
\begin{itemize}[leftmargin=*]
  \item well-capitalized (no additional capital constraint),
  \item continuously monitoring (one attempt every block),
  \item executed before other mempool flow in block mode.
\end{itemize}
This creates a harsh but realistic environment for LPs, as any significant mispricing that remains profitable
after funding costs is rapidly harvested.

\subsubsection{Liquidity Providers}
\label{subsec:LPs}

LPs deposit liquidity into the AMM and collect a pro-rata share of swap fees whenever their
positions are active. We model three LP cohorts:
(i) \emph{passive} LPs, who provide wide ranges and adjust infrequently;
(ii) \emph{active} LPs, who concentrate liquidity near the current price and manage positions more
aggressively; and
(iii) a MEV searcher performing \emph{Just-In-Time} (JIT) liquidity provision (the \emph{Jiter}).

\paragraph{LP state}
Each (non-JIT) LP $j$ maintains:
\begin{itemize}[leftmargin=*]
  \item a cash wallet $\mathcal{W}^1_{j,t}$ held entirely in token~1;
  \item a set of open positions composed by the lower tick $i_a$, upper tick $i_b$, liquidity deposited $L^{\mathrm{pos}}$ in $[i_a, i_b] \implies \mathcal{P}_{j,t}=\{(i_a,i_b,L^{\mathrm{pos}})_t\}$;
  \item internal review and cooldown clocks governing when the LP can revise positions; the review clock determines when the LP next reconsiders its range, while the cooldown clock enforces a waiting period after a burn before a new mint can be submitted;
  \item an associated rebalancing benchmark used for LVR accounting.
\end{itemize}
LPs are \emph{cash-budgeted}: between actions they do not carry token~0 in their wallet. When minting a new
position, the LP uses token~1 cash to acquire (conceptually, on the CEX) the exact token~0 amount required by
Uniswap~v3 for the chosen range at the prevailing CEX price $m_t$, and deposits the resulting pair of token
amounts into the AMM. When burning, the LP receives the position's underlying token amounts and accrued fees,
and then immediately converts any token~0 proceeds back into token~1 at price $m_t$. These conversions occur at the 
CEX price without explicit transaction fees but they induce permanent price impact.

\paragraph{Minting a position}
Consider a mint decision at block $t$ for LP $j$ with tick range $[i_a,i_b]$ (with $i_a<i_b$) and liquidity
$L>0$. Let
$
  s_t = \sqrt{P^{\mathrm{DEX}}_t},\qquad s_a=s(i_a),\qquad s_b=s(i_b),
$
where $s(\cdot)$ maps ticks to square-root prices (Section~\ref{subsec:univ3}). By the Uniswap~v3 algebra,
minting liquidity $L$ over $[s_a,s_b]$ requires depositing
$\bigl(\Delta x(L),\Delta y(L)\bigr)$, where
\begin{equation}
  \bigl(\Delta x(L),\Delta y(L)\bigr)
  =
  \begin{cases}
    \left(L\!\left(\dfrac{1}{s_a}-\dfrac{1}{s_b}\right),\,0\right), & s_t \le s_a,\\[1.0ex]
    \left(L\!\left(\dfrac{1}{s_t}-\dfrac{1}{s_b}\right),\,L(s_t-s_a)\right), & s_a < s_t < s_b,\\[1.0ex]
    \left(0,\,L(s_b-s_a)\right), & s_t \ge s_b.
  \end{cases}
  \label{eq:lp-mint-deposits}
\end{equation}
A key property is that, once $(s_t;s_a,s_b)$ is fixed, both $\Delta x(L)$ and $\Delta y(L)$ scale
linearly with $L$. Therefore, valuing token~0 at the CEX price $m_t$, the LP can
mint at maximum a liquidity $L^{\max}_{j,t}([i_a,i_b])$ such that
\begin{equation}
  \mathcal{W}^1_{j,t}
  =
  L^{\max}_{j,t}([i_a,i_b])\Bigl(m_t\,\Delta x(1) + \Delta y(1)\Bigr),
  \label{eq:lp-Lmax}
\end{equation}
where $\bigl(\Delta x(1),\Delta y(1)\bigr)$ denotes the deposit required for one unit of liquidity
in the same state $(s_t;s_a,s_b)$. Hence
\begin{equation}
  L^{\max}_{j,t}([i_a,i_b])
  =
  \frac{\mathcal{W}^1_{j,t}}{m_t\,\Delta x(1) + \Delta y(1)}.
  \label{eq:lp-Lmax-cash}
\end{equation}

To model heterogeneous aggressiveness while avoiding all the mints to be ``all-in'', each time the LP draws
a wallet utilization factor $\eta_{j,t}\in(0,1]$:
\begin{equation}
  Z_{j,t}\sim \log\mathcal{N}(\mu_{\mathrm{mint}},\sigma_{\mathrm{mint}}),\qquad
  \eta_{j,t}=\min\{1,Z_{j,t}\}.
  \label{eq:lp-utilization}
\end{equation}
The executed mint liquidity is
\begin{equation}
  \Delta L^{\text{exec}}_{j,t} \;=\; \eta_{j,t}\,L^{\max}_{j,t}([i_a,i_b]).
  \label{eq:lp-mint-executed}
\end{equation}
If $\Delta L^{\text{exec}}_{j,t}=0$, the mint is skipped. Otherwise, the LP uses token~1 cash to finance the mint: it
acquires on the CEX, at price $m_t$, the required token~0 amount $\Delta x(\Delta L^{\text{exec}}_{j,t})$, contributes the
token~1 amount $\Delta y(\Delta L^{\text{exec}}_{j,t})$ from its cash wallet, and deposits
$\bigl(\Delta x(\Delta L^{\text{exec}}_{j,t}),\Delta y(\Delta L^{\text{exec}}_{j,t})\bigr)$ into the AMM.
The cash wallet is debited by the token~1 value of this deposit:
\begin{equation}
  \mathcal{W}^1_{j,t^+}
  \;=\;
  \mathcal{W}^1_{j,t}
  -
  \Bigl(
    m_t\,\Delta x(\Delta L^{\text{exec}}_{j,t})
    +
    \Delta y(\Delta L^{\text{exec}}_{j,t})
  \Bigr),
  \label{eq:lp-wallet-debit-cash}
\end{equation}
and the position $(i_a,i_b,\Delta L^{\text{exec}}_{j,t})$ is added to $\mathcal{P}_{j,t}$. Any unspent budget
remains in token~1; in particular, the LP does not retain a residual token~0 inventory.

\paragraph{Burning a position}

When LP $j$ burns a position $p=(i_a,i_b,L^{\mathrm{pos}})\in\mathcal{P}_{j,t}$ at state $s_t$, the LP receives
(i) the position's current underlying token amounts $\left(\Delta y(L^{\mathrm{pos}}), \Delta x(L^{\mathrm{pos}})\right)$ and
(ii) its accrued fees $\left(F^{p}_{1,t}, F^{p}_{0,t}\right)$.
The underlying amounts are given by the same v3 formulas as \eqref{eq:lp-mint-deposits}, evaluated for the burned
position's range $[i_a,i_b]$ and liquidity $L=L^{\mathrm{pos}}$ at the current state $s_t$.
Under cash budgeting, token~0 proceeds are immediately converted into token~1 at price $m_t$, so the wallet
updates as
\begin{equation}
  \mathcal{W}^1_{j,t^+}
  \;=\;
  \mathcal{W}^1_{j,t}
  +
  \Delta y(L^{\mathrm{pos}})
  +
  F^{p}_{1,t}
  +
  m_t\Bigl(\Delta x(L^{\mathrm{pos}})+F^{p}_{0,t}\Bigr).
  \label{eq:lp-wallet-credit-cash}
\end{equation}

Fees are accrued tick-by-tick during swap execution: in each traversed tick
interval, the input-side fee flow is allocated to the LPs active in that interval proportionally to their
provided liquidity $L^{\mathrm{pos}}$.

\paragraph{Passive LPs}
Passive LPs provide baseline liquidity for swaps. Their strategy is:
\begin{itemize}[leftmargin=*]
  \item Wide ranges: Passive LPs mint positions symmetrically around the current DEX price, setting the range width as a fixed percentage of the entry price.
  \item Exogenous activity: Passive mint and burn decisions are probabilistic and do not depend on
  microstructure signals.
  \item Token-budgeted sizing: Conditional on minting a range, the position size is determined by
  \eqref{eq:lp-Lmax}--\eqref{eq:lp-mint-executed}, i.e.\ by available token inventories and the utilization
  factor.
\end{itemize}

\paragraph{Active LPs}
Active LPs provide more liquidity near the current price and manage positions dynamically. This active cohort is motivated by empirical evidence that Uniswap v3 LP outcomes depend strongly on range width and position timing, and by strategic-liquidity models in which LPs dynamically adjust ranges to optimize fee income and inventory exposure \cite{heimbach2022,fan2023}. They differ
from passive LPs along several dimensions:
\begin{itemize}[leftmargin=*]
  \item Adaptive width: The width $w_t$ (in ticks) of each active position is determined from an
  EWMA of a signal related to recent volatility, plus a binomial noise term to introduce heterogeneity:
  \begin{equation}
    w_t = \mathrm{min}\left(\mathrm{max}(\tilde w_t + \eta_t,w_{\min}),w_{\max}\right),
    \label{eq:active-width}
  \end{equation}
  where $\tilde w_t$ is a deterministic function of the chosen signal, $\eta_t$ is a random width
  perturbation, and $[w_{\min},w_{\max}]$ are bounds.
  \item Re-centering: Each active position tracks the number of consecutive blocks for which the
  DEX price has remained outside its range. If this counter exceeds a threshold
  $k_{\mathrm{out}} \sim U[k^{\min}_{\mathrm{out}},k^{\max}_{\mathrm{out}}]$, the LP enqueues a recenter intent:
  burn the old position and mint a new one centered at the current reference, with width $w_t$.
  The new mint is executed subject to token feasibility \eqref{eq:lp-Lmax}--\eqref{eq:lp-mint-executed}.
  \item Take-profit / stop-loss: Active LPs monitor the PnL of each position relative to a HODL
  benchmark of the initial tokens, meaning the value the LP would obtain by simply holding the two tokens that were initially required to mint the position and marking them to the external reference price. If the position's PnL exceeds a take-profit threshold
  $\theta_{\mathrm{TP}}$ or falls below a stop-loss threshold $-\theta_{\mathrm{SL}}$, the LP exits by burning
  the position.
\end{itemize}
Active LPs therefore tend to trade more frequently and run narrower ranges, making them more exposed to
LVR but potentially able to harvest higher fee density.

\paragraph{Review clocks and cooldowns}
Active and passive LPs are endowed with a review clock that governs how often they can revise positions.
Inter-review times are drawn from a geometric distribution with mean $\tau_{\mathrm{review}}$, so that at any
block roughly a fraction $1/\tau_{\mathrm{review}}$ of LPs are ``due'' to reconsider their ranges.
When an LP burns a position, it enters a cooldown during which it cannot mint new positions. This
mechanism introduces operational latency, mitigates high-frequency strategic oscillations and prevents
synchronization among all the LPs.

\paragraph{JIT LP (Jiter)}

The Jiter is a MEV-style liquidity provider, motivated by documented JIT liquidity and priority-ordering 
behavior in AMMs \cite{wan2022just,flashboys}, that tries to capture swap fees by providing liquidity only 
for the short window around a large swap. In each block, the Jiter is enabled only when $p_{\mathrm{JIT}}>0$, and $q^{\max}_{\mathrm{JIT}}>0$, where $p_{\mathrm{JIT}}$ is the Bernoulli attempt 
probability. Conditional on an attempt, the current implementation sorts pending swaps by input notional 
(valued in token~1) and targets only the single largest pending swap. When 
the selected target reaches execution, the swap is wrapped as jit mint $\to$ swap $\to$ jit burn: the Jiter 
mints a one-tick-wide position, lets the swap trade against this added liquidity, and then immediately burns 
the position. The mint size is not a fixed 90\% liquidity target. Instead, it is chosen by maximizing an 
expected one-tick profit approximation that trades off expected fee capture against the flash-financing cost 
of the minted principal. The parameter $q^{\max}_{\mathrm{JIT}}$ caps the Jiter's post-mint share of liquidity 
in the tick; in the baseline, $q^{\max}_{\mathrm{JIT}}=0.90$. The mint is skipped whenever the optimized size is 
non-positive or the expected fee capture does not cover the flash-loan fee. We model the mint as flash-funded and 
charge a per-mint financing cost given by the JIT flash-loan fee (0.01\%); upon burning, it withdraws principal 
plus accrued fees and nets any remaining token0 exposure by converting it back to token1 at the current reference 
price.


\paragraph{PnL Decomposition}
\label{paragraph:PnL_decomposition}
Let $F_t$ denote the cumulative fees earned by the LP up to time $t$, valued in token 1
using the CEX price $m_t$.

The key identity enforced by the model is
\begin{equation}
  V_t^{\mathrm{LP}} = V_t^{\mathrm{reb}} + F_t - \mathrm{LVR}_t,
  \label{eq:lvr-identity}
\end{equation}
where $\mathrm{LVR}_t$ is the cumulative loss-versus-rebalancing experienced by the LP
up to time $t$. Rearranging,
\begin{equation}
  \mathrm{LVR}_t = V_t^{\mathrm{reb}} - V_t^{\mathrm{LP}} + F_t.
\end{equation}

From \eqref{eq:lvr-identity} we obtain the following PnL notions:
\begin{itemize}[leftmargin=*]
  \item Unhedged PnL (full economic outcome):
        \begin{equation}
          \Pi_t^{\mathrm{unhedged}}
          = V_t^{\mathrm{LP}} - V_0^{\mathrm{LP}}.
        \end{equation}
        This includes all market beta and directional exposure.
  \item Hedged PnL (liquidity-provision economics only):
        \begin{equation}
          \Pi_t^{\mathrm{hedged}}
          = V_t^{\mathrm{LP}} - V_t^{\mathrm{reb}}
          = F_t - \mathrm{LVR}_t.
          \label{eq:hedged-pnl}
        \end{equation}
        This measures how much of the fees survive after subtracting adverse selection losses. In the case of the Jiter, $\Pi_t^{\mathrm{hedged}}$ is reduced also by the flash loan cost. 
\end{itemize}

By design, hedged PnL cannot be improved by taking more price risk: any strategy that alters
the LP's exposure path $x_t$ changes both the benchmark and the AMM wealth in lockstep.
Only the balance between fees collected and LVR determines $\Pi_t^{\mathrm{hedged}}$.

\subsubsection{Smart router}
To endogenize venue choice, we introduce a smart‑routing trader that allocates each potential trade between the automated market maker (AMM, i.e., the DEX) and the reference centralized exchange (CEX) according to a best‑execution criterion. The inclusion of a smart router is central to the model, as it captures the endogenous relative attractiveness of AMM versus CEX execution. When the liquidity‑taker fee on the AMM is high, the smart router optimally favors execution on the CEX, reducing the volume routed to the AMM and, consequently, the fee income accruing to liquidity providers (LPs). Conversely, when the liquidity‑taker fee is low, the smart router preferentially routes trades to the AMM, increasing fee revenues for LPs. This mechanism highlights how the ratio of CEX‑routed to AMM‑routed trades emerges as an informative metric of DEX efficiency

Smart router draws a token~1 notional
$
Y^{\mathrm{not}} \sim \exp(\mathcal{N}(\mu_{\mathrm{tr}},\sigma^2_{\mathrm{tr}}))
$
and a direction $\mathrm{side}\in\{X\!\to\!Y,\,Y\!\to\!X\}$ with equal probability.
For $X\!\to\!Y$ (sell token 0), the smart router fixes a token 0 input
$
\Delta x^{\mathrm{int}} = Y^{\mathrm{not}}/m_t
$
and queries the AMM quote $\Delta y^{\mathrm{DEX}}$.
The CEX benchmark output is $\Delta y^{\mathrm{CEX}} = \Delta x^{\mathrm{int}}\,m_t$
corresponding to executing the same sale at price $m_t$.
The best-execution requirement is
\begin{equation}
  \Delta y^{\mathrm{DEX}} \;\ge\; \Delta y^{\mathrm{CEX}},
  \label{eq:smart-bestexec-x2y}
\end{equation}
If \eqref{eq:smart-bestexec-x2y} fails, the trade
is routed entirely to the CEX; otherwise, the AMM leg is submitted to the mempool.
For $Y\!\to\!X$ (buy token~0), the trader fixes a token~1 input $\Delta y^{\mathrm{int}}=Y^{\mathrm{not}}$,
queries $\Delta x^{\mathrm{DEX}}$ and compares it to the CEX benchmark
\(
\Delta x^{\mathrm{CEX}} = \Delta y^{\mathrm{int}}/m_t
\)
via
\begin{equation}
  \Delta x^{\mathrm{DEX}} \;\ge\; \Delta x^{\mathrm{CEX}}.
  \label{eq:smart-bestexec-y2x}
\end{equation}
At execution time (when the mempool is replayed), the engine re-quotes the AMM against a baseline
computed from the last validated DEX reference price $P_t^{\mathrm{ref}}$.
The AMM execution is skipped if the realized quote violates a relative slippage tolerance, i.e.\ if
\(
\Delta y^{\mathrm{act}}/\Delta y^{\mathrm{base}} < 1-\delta_{\mathrm{slip}}
\)
(or analogously $\Delta x^{\mathrm{act}}/\Delta x^{\mathrm{base}} < 1-\delta_{\mathrm{slip}}$), where
$\delta_{\mathrm{slip}}\in(0,1)$ is configured.
This agent therefore provides a parsimonious micro-founded mechanism for fee-sensitive flow: as the AMM
fee increases or liquidity thins, a larger fraction of smart-router volume migrates to the CEX \footnote{It is worth noting, however, that in practice a trader may choose to route an order to an AMM not solely because it offers the best immediate price, but as part of a broader strategy involving other liquidity pools or DeFi protocols. Such behavior can be captured by relaxing the best-execution assumption and introducing a small tolerance around the optimal price.}. His PnL is evaluated according to the CEX price, hence his PnL will depend mainly on the discrepancies between our AMM and the reference market. The more the DEX is attractive, the higher the smart router's PnL will be. Finally, CEX trades experience permanent impact as explained in Section\ref{subsubsec:impact}.

\subsubsection{Noise Trader}



The noise trader represents uninformed or urgency-driven flow:
\begin{itemize}[leftmargin=*]
  \item Her trade sizes are drawn from a log-normal distribution in notional units of
        token 1, with parameters $(\mu_{\mathrm{trade}},\sigma_{\mathrm{trade}})$.
  \item Trade direction (buy vs.\ sell) is chosen at random with equal probability.
  \item Noise trader ignores the CEX price $m_t$ and thus have no valuation discipline.
  \item She enforces a slippage constraint: if the implied execution price on the DEX deviates
        from their reference quote by more than a tolerance (relative to the reference),
        the trade is aborted \footnote{This behaviour mimics the default slippage tolerance set using the Uniswap v3 web interface.}.
\end{itemize}

Noise traders pay fees to LPs and add volume, but their trades also move the AMM price away
from the CEX mid, creating opportunities for arbitrageurs and, hence, contributing to LVR. The noise trader's PnL is computed at the end of each block using the CEX price to convert everything in token 1.

\paragraph{Micro-step event probabilities.}
As already said in Section \ref{subsubsec:mempool}, the simulator advances in discrete blocks (indexed by $t=1,\dots,T$). Each block contains $B$ micro-steps (indexed by $\tau=1,\dots, B$) that represent within-block order arrival, while execution is aggregated at the block boundary via a mempool replay. For each event class $e$ (e.g., noise trades, LP mints, LP burns), the configuration specifies a micro-step arrival probability $p_e\in[0,1]$.

Within each block, arrivals are sampled independently across micro-steps as
\[
A_{e,t,\tau}\sim \text{Bernoulli}(p_e),\qquad \tau=1,\dots,B,
\]
and one event of type $e$ is instantiated whenever $A_{e,t,\tau}=1$. The resulting number of arrivals within block $t$ is
\[
N_{e,t}=\sum_{\tau=1}^B A_{e,t,\tau}\sim \text{Binomial}(B,p_e),
\]
hence $\mathbb{E}[N_{e,t}]=Bp_e$ and the expected number of events per block scales linearly with the block time $B$.

Each realized event is instantiated as an intent and assigned to an eligible agent (or position) drawn from the corresponding candidate set (e.g., agents that are “due” and not in cooldown), allowing multiple intents per agent in a block when applicable. All intents accumulated over the block are then executed at the block boundary (mempool replay), which decouples stochastic arrival from on-chain ordering while keeping the arrival process directly controllable through $p_e$.

\subsection{Dynamic Fee Schedules}
\label{subsec:fees}

A central feature of the model is a time-varying taker fee $f_t \in [0,1)$, representing the fraction of 
notional charged to traders and paid to LPs.
At each step, a raw fee proposal $f^{\mathrm{raw}}_t$ is computed from a chosen signal and then passed 
through a common controller that bounds the fee and avoid too frequent updates. In market-design terms, this controller treats the AMM fee as an endogenous transaction-cost rule, analogous 
to fee and tax interventions studied in artificial financial markets, but here targeted at AMM adverse 
selection \cite{westerhoff2006,mannaro2008,pellizzari2009,CARTEA2025105134}.

We consider three modes, based on:
\begin{enumerate}[label=\roman*), leftmargin=*]
  \item static
  \item volatility
  \item toxicity
\end{enumerate}

\paragraph{EWMA smoothing}
Several controllers use a smoothed version of a noisy observation process $x_t$ through an exponentially 
weighted moving average (EWMA)
\begin{equation}
  v_t = \lambda\, v_{t-1} + (1-\lambda)\, x_t,
  \label{eq:ewma}
\end{equation}
with $\lambda\in(0,1)$. 

\paragraph{Static fee}
The fee is constant:
\begin{equation}
  f_t = f_0, \qquad \forall t,
  \label{eq:static-fee}
\end{equation}
where $f_0$ is the initial fee level (and the fixed fee in static mode).

\paragraph{Volatility based fee}
This controller reacts to realized volatility of the price. We employ two variants according to the venue we take as reference. For simplicity, we report here the specification relative to the CEX price. It is totally similar in the case of the (validated) DEX price\footnote{Using the CEX price requires the AMM to have an oracle able to read that price. In practice, this is possible but it adds a new risk relative to oracle failures.}.
Define 
\begin{equation}
  vol\_obs_t = \bigl(\ln(m_t) - \ln(m_{t-1})\bigr)^2,
\end{equation}
and its EWMA $\hat{\sigma}_t = \mathrm{EWMA}(vol\_obs_t)$ as in \eqref{eq:ewma}.
The raw fee proposal is:
\begin{equation}
  f^{\mathrm{raw}}_t = k_{\sigma}\, \sqrt{\hat{\sigma}_t},
  \label{eq:vol-fee}
\end{equation}
where $k_{\sigma}>0$ scales sensitivity.

\paragraph{Toxicity based fee}
This controller measures arbitrage ``toxicity'' via the price gap that exceeds the current fee band.
Let $f_{\mathrm{current}}$ be the fee currently applied by the pool and define the fee band in log-space:
\begin{equation}
  f_{\mathrm{band ln}} = -\ln(1-f_{\mathrm{current}}).
\end{equation}
Let the absolute log price gap be
\begin{equation}
      \Delta_{\mathrm{cex\_dex}} = \bigl|\ln(P^{\mathrm{DEX}}_t) - \ln(m_t)\bigr|.
\end{equation}
The excess basis observation (using the \emph{current} fee) is
\begin{equation}
  B_{obs,t} = \max\!\bigl(0,\ \Delta_{\mathrm{cex\_dex}} - f_{\mathrm{band ln}}\bigr),
  \label{eq:basis-obs}
\end{equation}
which is then smoothed as $B_{\hat{t}}=\mathrm{EWMA}(B_{obs,t})$.
Converting to ticks using the log tick size $\Delta i\log{\gamma}$,
\begin{equation}
  \tilde{B}_t = \frac{B_{\hat{t}}}{\Delta i\log{\gamma}}.
\end{equation}
The raw fee is:
\begin{equation}
  f^{\mathrm{raw}}_t = k_{\mathrm{tox}}\, \tilde{B}_t ,
  \label{eq:tox-fee}
\end{equation}
with $k_{\mathrm{tox}}>0$.

In the volatility and toxicity modes, $f_0$ is the initial fee and the static-fee level. 
The implemented dynamic proposal has the form
\begin{equation}
  f_t^{\mathrm{raw}}=k\,s_t,
\end{equation}
where $s_t=\sqrt{\hat\sigma_t}$ for volatility controllers and $s_t=\tilde B_t$ for the toxicity controller. The proposal is then subject to the common step-size controller and clamped to $[f_{\min},f_{\max}]$; hence the dynamic floor is $f_{\min}$.


\paragraph{Shared fee update controller}
Regardless of mode, $f^{\mathrm{raw}}_t$ is transformed into the applied fee through the same mechanism:

(i) Define the increment
\begin{equation}
  \Delta f_t = f^{\mathrm{raw}}_t - f_t
\end{equation}

(ii) Hysteresis and step-size limits \newline
If the change is too small, no update is performed. If it is too high, it is capped. Let $\Delta f_{max}, \Delta f_{min}$ be the max and min possible fee updates:
\begin{align}
  |\Delta f_t| < \Delta f_{min} 
  \quad&\Rightarrow\quad \text{no update.} \\ 
  |\Delta f_t| > \Delta f_{max} \quad&\Rightarrow\quad \Delta f_t=\Delta f_{max}
\end{align}

(iii) Fee update \newline
The new fee is set to

\begin{equation}
    f_{t+1} = f_t + \Delta f_t
\end{equation}

clamped between a minimum and maximum fee value to prevent extreme situations.

(iv) Cooldown \newline 
After a fee change is scheduled, further changes are ignored for $\tau_{\text{cool}}$ steps, preventing rapid oscillations.
Overall, this controller keeps fees bounded, smooth, and responsive only to meaningful shifts in the underlying signal.

This controller ensures that the fee path is bounded, piecewise smooth, and only reacts to material changes in the underlying signals.

\subsection{Initialization}
The model is initialized by setting an initial pool price $P^{\mathrm{DEX}}_0$ and seeding the tick grid with a baseline liquidity profile, distributed as a binomial mass concentrated around the initial DEX price. This “liquidity hill” is allowed to evolve endogenously as swaps move the active price across ticks, but the positions that constitute it are assigned to a separate LP cohort that is excluded from the PnL accounting. Its sole purpose is to provide initial depth so that early trades can execute without pathological price jumps, thereby ensuring a well-defined transient phase before the endogenous LP population and trading flows drive the system toward its stationary regime.

\section{Results}
\label{sec:results}

In this section we illustrate the results obtained with the ABM framework introduced in the previous sections. The goal 
is to assess whether dynamic fees can mitigate adverse selection (LVR) under block-time execution, and to quantify the 
trade-offs with competitiveness (DEX share) and with strategic liquidity provision (active LPs and JIT liquidity).

We consider three models with increased complexity:
\begin{enumerate}
    \item \textbf{Model 0:} arbitrageur, noise trader, smart router, and a passive LP cohort only. This baseline isolates 
    block-time stale-quote risk without strategic liquidity provision.
    \item \textbf{Model 1:} we introduce an active (more concentrated) LP cohort alongside passive LPs (equal split of 
    the LP population) and compare hedged PnL across cohorts under different fee policies.
    \item \textbf{Model 2:} we additionally allow a JIT liquidity provider (the ``Jiter'') to operate, to study how 
    MEV-style liquidity affects the incidence of fees and LVR across LP cohorts.
\end{enumerate}

All simulations share a common set of parameters (Table \ref{tab:common_params}). As stated in the introduction, 
our objective is not to construct a data-generating process that accurately reproduces empirical market properties, 
but rather to develop a framework that allows us to study how different types of agents interact and how the 
introduction of dynamic fee mechanisms affects the economics of liquidity provision. Accordingly, parameter values 
are chosen to generate plausible and interpretable dynamics, providing a meaningful setting in which qualitative 
reasoning and comparative analysis can be carried out. Several parameters (e.g., the order-flow intensity per block) 
are chosen to be consistent with the empirical calibration in \cite{dinosse_gatta}. We anyway provides a brief
section in which we shows that even without a proper empirical calibration, the model is able to generate realistic 
price and trade dynamics, see Section \ref{sec:stylized}.

Unless otherwise stated, summary figures report mean \(\pm\) standard error across 100 independent seeds per 
\((\text{model},\text{fee schedule})\) configuration.

The main metrics of interest are:
\begin{itemize}
    \item the \textbf{hedged LP PnL} (Section~\ref{paragraph:PnL_decomposition}), which quantifies the profitability of 
    liquidity provision in the AMM net of market risk:
    \begin{equation}
        \Pi_T^{\mathrm{hedged}} = F_T - \mathrm{LVR}_T - \mathbf{1}_{JIT}C_{\mathrm{flash}},
    \end{equation}
    where $\mathbf{1}_{JIT}=1$ for the Jiter cohort and $0$ otherwise;
    \item the \textbf{fraction of trades routed to the DEX} relative to total trades (DEX share), which serves as a 
    measure of the AMM’s attractiveness compared to the CEX:
    \begin{equation}
        \text{DEX}_{share} = \frac{\text{\#DEX trades }}{\text{\# CEX+DEX trades}}
    \end{equation}
    \item the \textbf{liquidity-taker fee level} $f_t$. Holding all other parameters fixed, higher fees increase LP 
    revenues but reduce the competitiveness of the AMM, thereby discouraging order routing by the smart router in favor 
    of the CEX.
\end{itemize}

\begingroup
\scriptsize
\renewcommand{\arraystretch}{1.05}
\begin{longtable}{|p{0.19\textwidth}|p{0.56\textwidth}|p{0.19\textwidth}|}
\caption{Baseline simulation parameters shared across all models.}
\label{tab:common_params}\\
\hline
\textbf{Symbol} & \textbf{Description} & \textbf{Value} \\
\hline\hline

\rowcolor{groupgray}\multicolumn{3}{|l|}{\textbf{Timeline \& horizon}}\\
\hline
\(\tau\) & Block duration (mempool length), in seconds / micro-steps per block & \(5\) \\
\hline
\(T\) & Total number of simulation blocks & \(13 000\) \\
\hline
\(T_{\text{skip}}\) & Blocks skipped to reach equilibrium & \(3 000\) \\
\hline\hline

\rowcolor{groupgray}\multicolumn{3}{|l|}{\textbf{Reference market (CEX): Heston}}\\
\hline
\(\mu\) & Drift of \(\ln m_t\) per micro-step & \(0\) \\
\hline
\(\sigma_0\) & Initial instantaneous volatility per micro-step (\(v_0=\sigma_0^2\)) & \(1.5\times10^{-4}\) \\
\hline
\(\kappa\) & Mean reversion speed of variance \(v_t\) & \(1.0\) \\
\hline
\(\theta\) & Long-run variance level & \(1.0\times10^{-8}\) \\
\hline
\(\sigma_v\) & Volatility of variance (vol-of-vol) & \(1.0\times10^{-3}\) \\
\hline
\(\rho\) & Corr.\ between price and variance shocks & \(-0.5\) \\
\hline
\(v_0\) & Initial variance level & \(2.25\times10^{-8}\) \\
\hline\hline

\rowcolor{groupgray}\multicolumn{3}{|l|}{\textbf{Initial prices and CEX impact}}\\
\hline
\(P^{\mathrm{DEX}}_0,\,m_0\) & Initial DEX and CEX prices (token~1 per token~0) & \((2000,\,2000)\) \\
\hline
\(\eta_{\mathrm{imp}},\,\xi\) & Permanent CEX-impact coefficient and exponent in \(m_{\tau^+}=\max\{10^{-12},m_{\tau^-}+\eta_{\mathrm{imp}}\operatorname{sign}(\Delta a)|\Delta a|^\xi\}\) & \((1.0\times10^{-3},\,0.5)\) \\
\hline\hline

\rowcolor{groupgray}\multicolumn{3}{|l|}{\textbf{Liquidity taker arrivals and trade sizes}}\\
\hline
\(\lambda_{\mathrm{smart}},\,\lambda_{\mathrm{noise}}\) & Poisson arrival intensities per micro-step / second & \((0.16,\,0.16)\) \\
\hline
\(\lambda_{\mathrm{act\ mint}},\lambda_{\mathrm{pass\ mint}},\lambda_{\mathrm{pass\ burn}}\) & LP-intent Poisson intensities per micro-step / second & \((0.10,\,0.10,\,0.10)\) \\
\hline
\(\mu_{\mathrm{tr}},\,\sigma_{\mathrm{tr}}\) & Log-normal location and scale for both smart-router and noise-trader token-1 notional & \((2.5,\,1.5)\) \\
\hline
\(\delta_{\text{slip}}\) & Max tolerated relative slippage (trade aborted if exceeded) & \(0.01\) \\
\hline\hline

\rowcolor{groupgray}\multicolumn{3}{|l|}{\textbf{Arbitrageurs}}\\
\hline
\(\phi_{\text{flash}}\) & Flash-loan funding cost fraction & \(1.0\times10^{-4}\) \\
\hline\hline

\rowcolor{groupgray}\multicolumn{3}{|l|}{\textbf{Liquidity providers (LPs)}}\\
\hline
\(N_{\text{LP}}\) & Number of strategic LP agents & \(100\) \\
\hline
\(w^{\text{pass}}_{\%}\) & Passive LP total width (\(\pm w^{\text{pass}}_{\%}/2\) around reference price) & \(5\%\) \\
\hline
\(\tau_{\text{review}}\) & Mean LP review cadence (seconds / micro-steps) & \(25\) \\
\hline
\(w_{\min}\) & Minimum programmatic mint width (ticks) & \(10\) \\
\hline
\(w_{\max}\) & Maximum programmatic mint width (ticks) & \(1.77454\times10^{6}\) \\
\hline
\(h_{\text{width}}\) & Half-life of EWMA absolute CEX log-return driving active-LP widths & \(1\) \\
\hline
\(s\) & Width sensitivity to the tick-scaled EWMA absolute log-return & \(1\) \\
\hline
\(n_{\text{bin}},\,p_{\text{bin}}\) & Binomial noise on widths: trials and success probability & \((70,\,0.5)\) \\
\hline
\(k^{\min}_{\text{out}},\,k^{\max}_{\text{out}}\) & Out-of-range blocks before recenter threshold, \(k_{\text{out}}\!\sim\!U[\cdot]\) & \((10,\,20)\) \\
\hline
\(\mu_{\text{mint}},\,\sigma_{\text{mint}}\) & Log-normal parameters for mint-size scaling & \((-1,\,1.5)\) \\
\hline
\(\theta_{\text{TP}}\) & Take-profit threshold (relative to initial HODL value) & \(0.10\) \\
\hline
\(\theta_{\text{SL}}\) & Stop-loss threshold (relative to initial HODL value) & \(0.20\) \\
\hline
\(c_{\mathrm{cool}}\) & Cooldown after a burn before the same LP can mint again; drawn uniformly and measured in review-clock units & \(U\{3,\ldots,8\}\) blocks (\(5\times U\{3,\ldots,8\}\) seconds in the baseline) \\
\hline
\(\mathcal W^1_{j,0}\) & Initial strategic LP token-1 wallet; seed LPs and Jiter start with zero wallet cash & \(V^{\mathrm{seed}}_0/N_{\mathrm{LP}}\) \\
\hline\hline

\rowcolor{groupgray}\multicolumn{3}{|l|}{\textbf{JIT liquidity provider (MEV searcher)}}\\
\hline
\(p_{\mathrm{JIT}}\) & Bernoulli JIT attempt probability per block & \(0\) in Models~0--1; \(1\) in Model~2 \\
\hline
\(q^{\max}_{\text{JIT}}\) & Cap on post-mint JIT share of targeted-tick liquidity & \(0.90\) \\
\hline\hline

\rowcolor{groupgray}\multicolumn{3}{|l|}{\textbf{Initialization}}\\
\hline
\(N^{(0)}_{\text{bin}}\) & Binomial hill depth used to seed initial liquidity distribution & \(450\) \\
\hline
\(L^{(0)}_{\text{tot}}\) & Total initial liquidity deployed across seed LPs & \(5.0\times10^{5}\) \\
\hline\hline

\rowcolor{groupgray}\multicolumn{3}{|l|}{\textbf{Fee controller}}\\
\hline
\(f_0\) & Initial taker fee and static-mode fee; not the dynamic floor & \(1.0\times10^{-4}\) \\
\hline
\(f_{\min},\,f_{\max}\) & Fee bounds enforced by controller & \((1.0\times10^{-5},\,5.0\times10^{-2})\) \\
\hline
\(h_{\text{fee}}\) & EWMA half-life for controller signals; \(\lambda=\exp[-\log(2)/h_{\text{fee}}]\) & \(2\) \\
\hline
\(k_{\sigma}\) & Volatility-to-fee multiplier & \(1.0\) \\
\hline
\(k_{\text{basis}}\) & Toxicity-to-fee multiplier & \(5.0\times10^{-4}\) \\
\hline
\(\Delta f_{\min}\) & Min absolute fee change to commit & \(5.0\times10^{-8}\) \\
\hline
\(\Delta f_{\max}\) & Max fee change per update step & \(0.01\) \\
\hline
\(\tau_{\text{cool}}\) & Fee update cooldown (blocks) & \(0\) \\
\hline
\end{longtable}
\endgroup

\FloatBarrier
\subsection{Microstructure diagnostics}
\label{sec:stylized}
Before comparing fee policies, we validate that the ABM reproduces the expected microstructure channel: block-time 
latency generates transient CEX--DEX deviations that are harvested by priority arbitrage, implying adverse selection (LVR) for LPs.

\begin{figure}[!htbp]
  \centering
  \begin{subfigure}{0.95\linewidth}
    \centering
    \includegraphics[width=\linewidth]{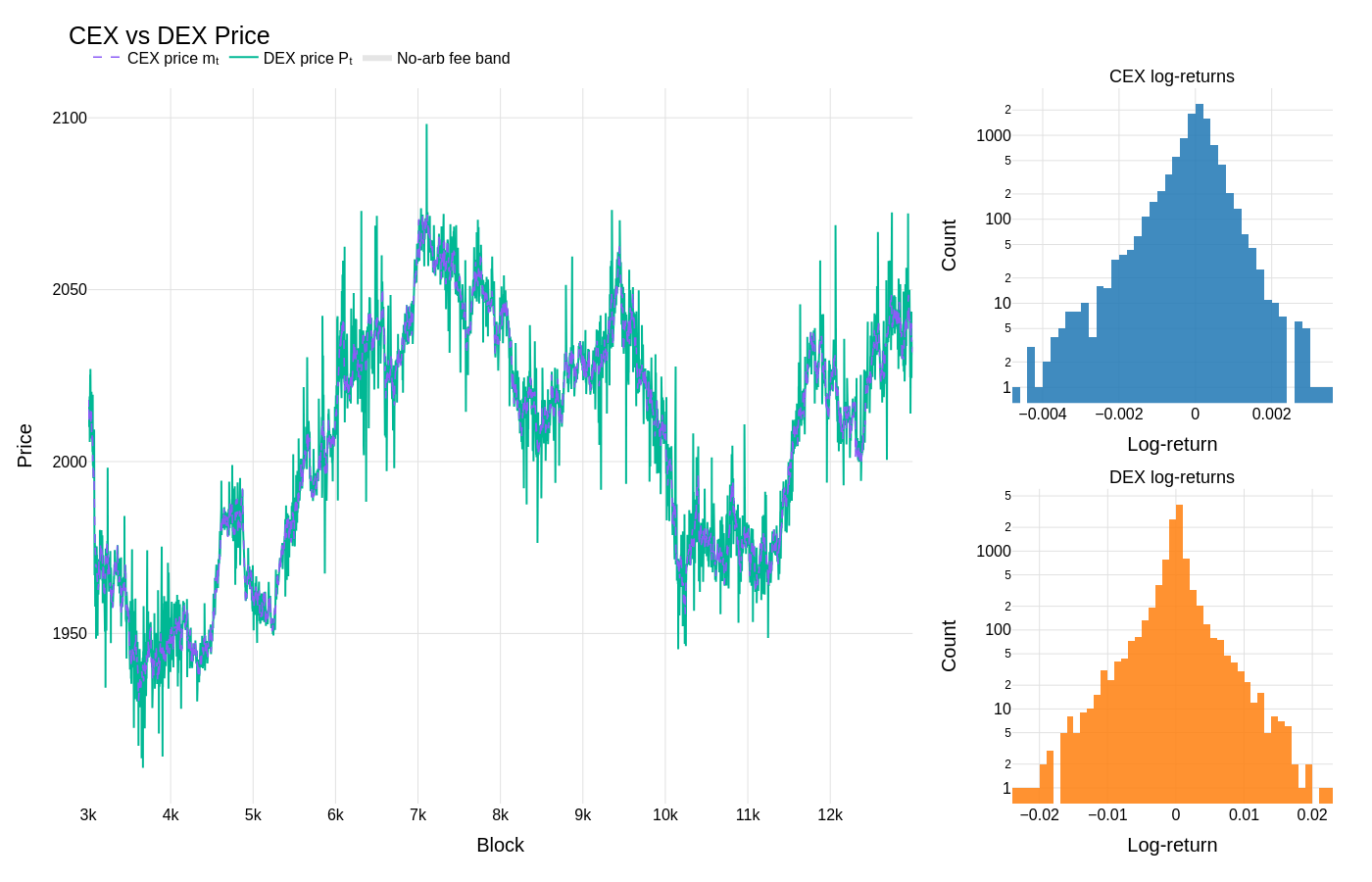}
    \caption{End-of-block CEX and DEX prices with the no-arbitrage band (representative static-fee run).}
  \end{subfigure}

  \vspace{0.4em}

  \begin{subfigure}{0.95\linewidth}
    \centering
    \includegraphics[width=\linewidth]{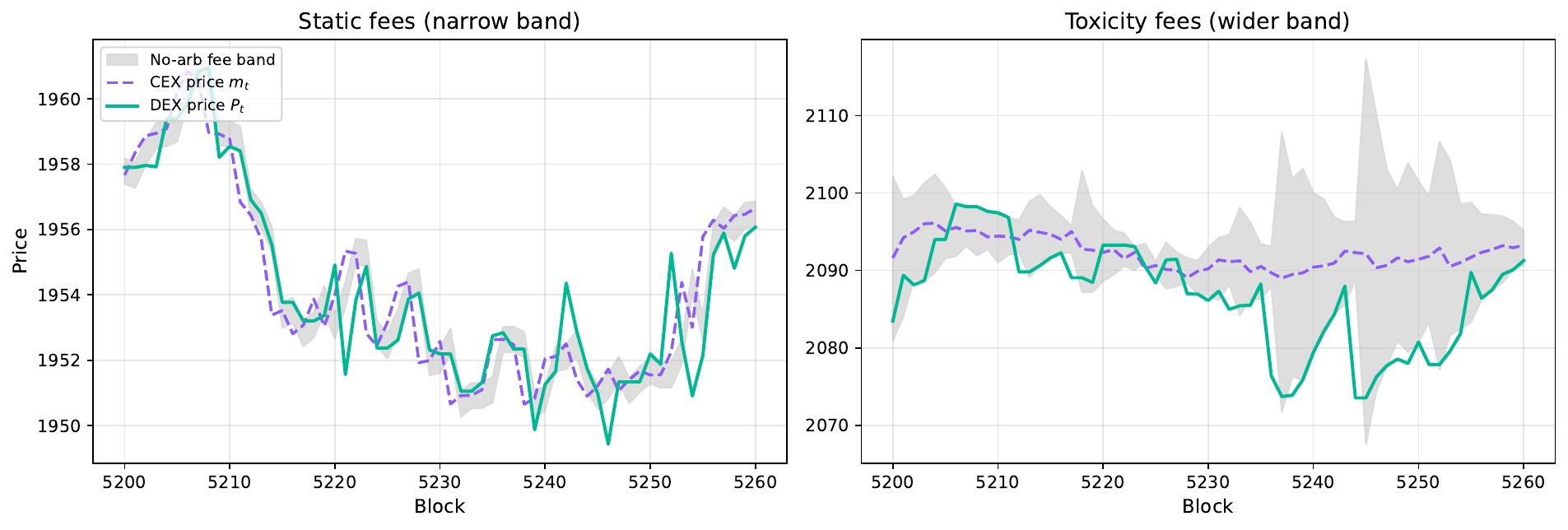}
    \caption{Zoomed comparison of the same Model~0 seed and parameters under static fees (left) and toxicity-based fees (right).}
  \end{subfigure}
  \caption{Microstructure illustration in Model~0. Panel (a) shows the static-fee CEX and DEX prices with the no-arbitrage band. 
  Panel (b) shows that toxicity-based fees widen the no-arbitrage band during stale-quote episodes, reducing executed arbitrage 
  and attenuating the next-block correction of the DEX price toward the CEX price.}
  \label{fig:microstructure_prices}
\end{figure}

Figure~\ref{fig:microstructure_prices} compares the reference (CEX) price with the AMM (DEX) price and the implied 
no-arbitrage band. The DEX series exhibits pronounced spikes because the CEX price evolves within each block while 
on-chain swaps are executed at the block boundary. More broadly, this channel is consistent with artificial-market evidence that latency and high-frequency priority mechanisms can generate short-horizon volatility and instability even when they improve price alignment in normal periods \cite{leal2016,gao2024,yagi2020}. Under static fees, mispricings are then corrected by arbitrage at
the beginning of the next block, generating the spike-and-reversion dynamics documented empirically in \cite{dinosse_gatta}. 
Under toxicity-based fees, the same stale-quote gaps are often absorbed by a wider no-arbitrage band, so fewer 
economically sensible arbitrage swaps are executed and the visible pull of the DEX price back toward the CEX price is attenuated. 
In the representative Model~0 seed used for the zoom, executed arbitrage swaps fall from 9,725 under static fees to 
6,571 under toxicity fees, while in-band arbitrage no-ops rise from 3,275 to 6,428. Sparse or low-liquidity tick regions 
can also stop an arbitrage swap before its unconstrained target; this liquidity-friction mechanism is present under both 
fee schedules, whereas the comparative attenuation in panel (b) is driven by the wider dynamic-fee no-arbitrage wedge. 
The apparent lag between the CEX series and the shaded band is structural: the band is computed from the lagged snapshot 
\(m_{t-1}\), whereas the plotted \(m_t\) is the contemporaneous end-of-block reference price after intra-block diffusion.

\begin{figure}[!htbp]
  \centering
  \includegraphics[width=0.95\linewidth]{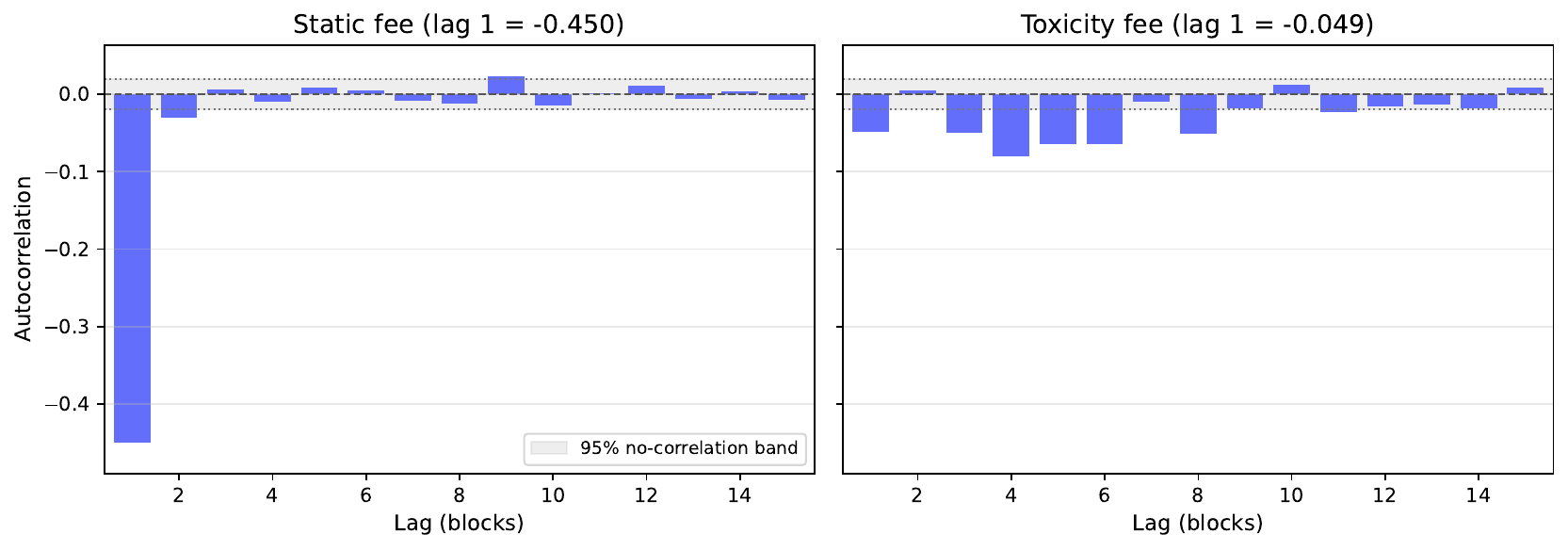}
  \caption{Autocorrelation function of end-of-block DEX log returns in the representative Model~0 seed used in Figure~\ref{fig:microstructure_prices}. 
  Static fees generate a pronounced lag-1 negative spike (\(-0.450\)); toxicity-based fees attenuate the lag-1 autocorrelation 
  to \(-0.049\), consistent with fewer next-block arbitrage corrections.}
  \label{fig:microstructure_acf}
\end{figure}

Another informative diagnostic is the autocorrelation function (ACF) of end-of-block DEX returns. 
Figure~\ref{fig:microstructure_acf} contrasts the static-fee baseline with the toxicity controller. Under static fees, 
the ACF shows a pronounced negative spike at lag 1, indicating short-horizon mean reversion across consecutive blocks. 
A qualitatively similar lag-1 anticorrelation has been documented in \cite{dinosse_gatta} and is often linked to arbitrage dynamics; 
our simulations show that block-time non-atomic CEX--DEX arbitrage alone can generate this signature. Under toxicity-based fees, 
the lag-1 spike is strongly attenuated in the same representative seed, consistent with the lower rate of economically 
sensible arbitrage corrections documented in Figure~\ref{fig:microstructure_prices}.

\FloatBarrier
\subsection{Cross-scenario outcome summary}
\label{subsec:results_summary}
We now summarize outcomes across all \((\text{model},\text{fee schedule})\) combinations using aggregated cross-scenario 
figures. This presentation focuses on the paper's central 
question: whether dynamic fees can offset block-time adverse selection without rendering the AMM uncompetitive, and how 
the answer changes once liquidity provision becomes more strategic.

\paragraph{Hedged PnL across models and fee modes.}

\begin{figure}[!htbp]
  \centering
  \includegraphics[width=\linewidth]{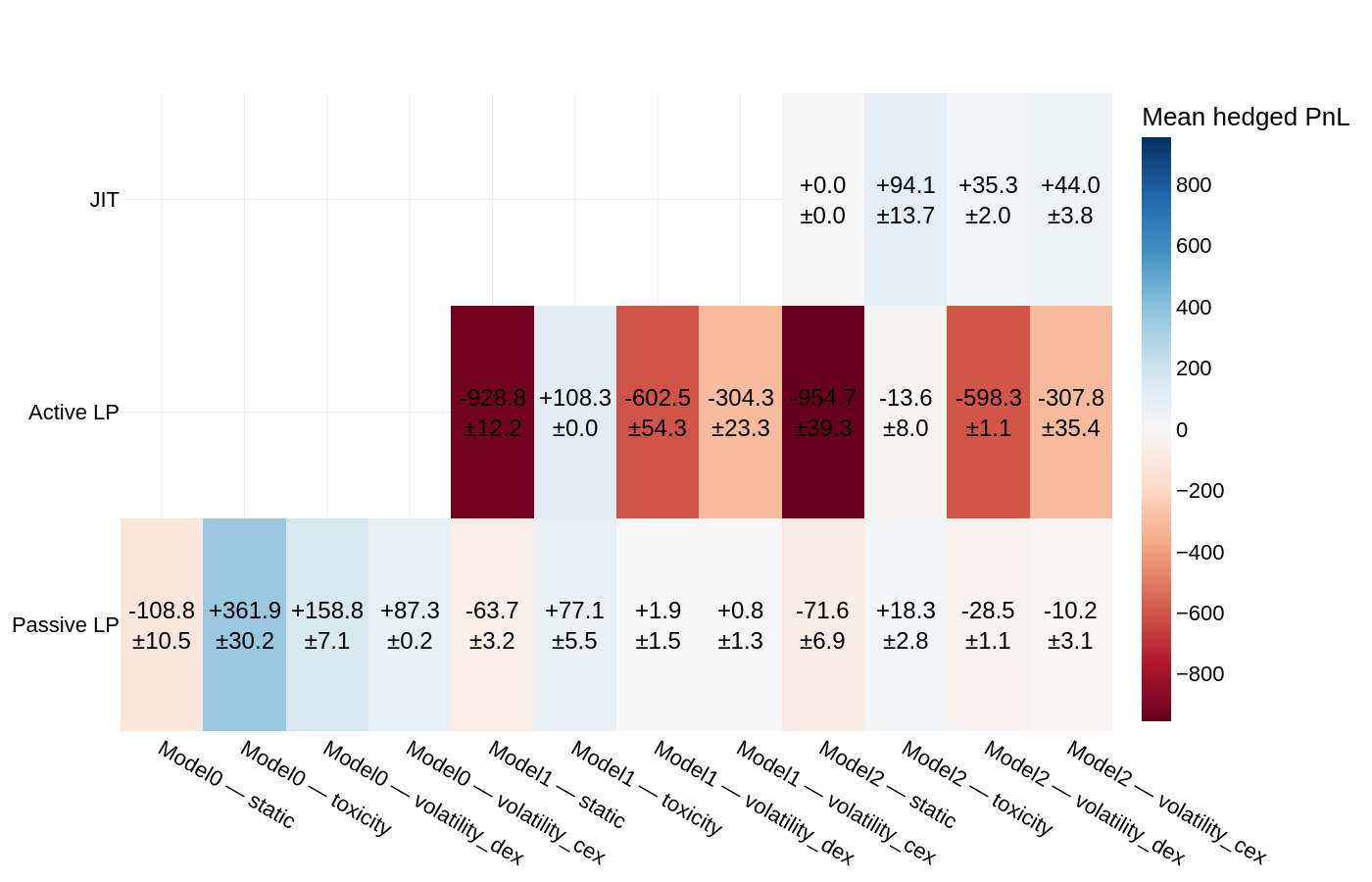}
  \caption{Final hedged PnL (token-1) across models and fee modes. Cells report mean \(\pm\) standard error across 100 seeds. 
  Inactive cohorts (e.g., JIT in Models 0--1) are left blank.}
  \label{fig:pnl_heatmap}
\end{figure}

Figure~\ref{fig:pnl_heatmap} reports the final hedged PnL by cohort (token-1 numéraire). The static-fee benchmark remains
unfavorable for standing liquidity in every model. Passive LP hedged PnL is negative in Model~0 (\(-108.8\pm 10.5\)),
Model~1 (\(-63.7\pm 3.2\)), and Model~2 (\(-71.6\pm 6.9\)); when active LPs are present, they are substantially more
exposed to adverse selection under static fees, with hedged PnL of \(-928.8\pm 12.2\) in Model~1 and
\(-954.7\pm 39.3\) in Model~2. This establishes the baseline microstructure result at the aggregate level: a fixed fee
does not compensate stale-quote losses, and concentration amplifies that exposure.

Dynamic fees materially change these outcomes, but their incidence depends on both the fee signal and the liquidity cohort.
In Model~0, every dynamic controller turns passive LP hedged PnL positive: toxicity delivers \(361.9\pm 30.2\),
DEX-volatility delivers \(158.8\pm 7.1\), and CEX-volatility delivers \(87.3\pm 0.2\). Once active liquidity is introduced
in Model~1, toxicity is the only rule that makes both standing LP cohorts positive, with \(77.1\pm 5.5\) for passive LPs
and \(108.3\pm 0.0\) for active LPs. By contrast, the volatility-based rules leave passive LPs approximately break-even
or mildly positive but keep active LPs negative (\(-602.5\pm 54.3\) under DEX volatility and \(-304.3\pm 23.3\) under
CEX volatility), indicating that generic volatility signals do not compensate concentrated liquidity for the realized
stale-quote losses in this configuration.

In Model~2, where JIT liquidity is admitted, the toxicity controller no longer protects all standing liquidity. It leaves
passive LPs slightly positive (\(18.3\pm 2.8\)) but active LPs slightly negative (\(-13.6\pm 8.0\)); the two volatility-based
rules leave both standing cohorts negative or close to zero. The Jiter is neutral under static fees
(\(0.0\pm 0.0\)) and profitable under dynamic fees, with the largest hedged PnL under toxicity (\(94.1\pm 13.7\), compared
with \(35.3\pm 2.0\) under DEX volatility and \(44.0\pm 3.8\) under CEX volatility). The heatmap therefore supports a more
qualified interpretation: dynamic fees can compensate standing liquidity for LVR by increasing fee income, especially
when based on the toxicity signal, but the introduction of event-contingent JIT liquidity redirects part of the enlarged
fee pool and prevents a uniform improvement for active LPs.

\paragraph{Competitiveness: DEX share.}
The smart router trades off execution quality against fees, so higher endogenous fees can reduce AMM competitiveness. 
Table~\ref{tab:routing_fee_summary} reports three routing-and-fee diagnostics used to evaluate this trade-off: 
the smart-router DEX share, the cumulative fee value earned by standing LP cohorts, and the time-averaged taker fee. 
The first diagnostic shows a systematic competitiveness cost. Static fees yield the highest DEX share in every model 
(\(35.50\%\) in Model~0, \(39.11\%\) in Model~1, and \(38.53\%\) in Model~2), whereas dynamic controllers lower routed 
DEX share to roughly \(22.77\%\)--\(27.94\%\). The toxicity rule produces the largest decline in routed flow, especially 
in Model~0, consistent with its higher average fee level. The volatility-based rules preserve slightly more order flow 
than toxicity, but they still remain well below the static benchmark. Importantly, the AMM remains economically relevant 
in all configurations; dynamic fees reduce DEX share rather than eliminating routed flow. The improvement in LP protection 
documented in Figure~\ref{fig:pnl_heatmap} is therefore achieved through a measurable, but partial, loss of competitiveness 
rather than through a collapse in trading activity. It is worth noting, however, that the tracked DEX share should be seen 
as a lower bound for the AMM's competitiveness, since the smart router is a simplified proxy for real-world routing logic 
and may not capture all the nuances of how traders choose between venues in practice. For instance, a trader can choose to 
swap on the DEX for other reasons beyond pure price and fee considerations, such as privacy, censorship resistance, or 
simply as a part of a bigger strategy that involves on-chain interactions.

\paragraph{Fee levels and fee incidence across cohorts.}
The remaining columns of Table~\ref{tab:routing_fee_summary} report the time-averaged taker fee and the corresponding 
fee-incidence decomposition across passive and active standing LP cohorts. Toxicity is systematically the highest-fee 
dynamic regime, reaching \(14.48\) bps in Model~0 and about \(11\)--\(12\) bps in Models 1 and Model 2. The DEX-volatility controller 
is intermediate, at \(7.25\) bps in Model~0 and about \(4.40\)--\(5.22\) bps thereafter, while the CEX-volatility controller 
remains close to \(5.3\) bps across all models. In Model~0, where all standing liquidity is passive, higher fees map almost 
one-for-one into passive fee income: cumulative standing-LP fee value rises from \(54.7\pm0.2\) under static fees to 
\(532.0\pm4.1\) under toxicity.

Once passive and active liquidity coexist, the fee distribution becomes highly asymmetric. In Models~1 and~2, the active 
cohort receives the dominant share of dynamic fee income, especially under toxicity, where cumulative standing-LP fee value 
rises to \(1524.1\pm92.2\) in Model~1 and \(1303.8\pm9.2\) in Model~2. This comparison is important because it shows that
large fee revenues do not, by themselves, establish that adverse-selection losses have fallen. Under both volatility-based
rules, active LPs collect substantial fees yet remain strongly negative in hedged PnL terms (Figure~\ref{fig:pnl_heatmap}).
What matters is not only charging more, but charging more precisely in states associated with stale-quote risk. In Model~1,
toxicity is the only specification that simultaneously raises the fee wedge and moves active LP hedged PnL into positive
territory; in Model~2, after JIT liquidity is admitted, the same controller leaves passive LPs slightly positive but active
LPs approximately break-even or mildly negative. The aggregate message is therefore two-sided: dynamic fees can compensate
LPs for LVR through additional fee income, but the signal design and the presence of strategic, latency-sensitive liquidity
determine who receives that compensation.

\begin{table}[!htbp]
  \centering
  \scriptsize
  \renewcommand{\arraystretch}{1.12}
  \caption{Routing and fee outcomes across model specifications and fee schedules. The reported diagnostics are the smart-router DEX share, cumulative fee value, and time-averaged taker fee. Entries are means \(\pm\) standard errors across 100 independent seeds. DEX share is measured as the percentage of smart-router trades routed to the AMM; passive, active, and total fee values are cumulative token-1 fee revenues for the displayed standing LP cohorts, with total equal to passive plus active fees; average taker fees are reported in basis points. Static rows use the fixed baseline fee \(f_0=1\) bp.}
  \label{tab:routing_fee_summary}
  \resizebox{\textwidth}{!}{%
  \begin{tabular}{llccccc}
    \toprule
    \textbf{Model} & \textbf{Fee mode} & \textbf{DEX share (\%)} & \textbf{Passive fees} & \textbf{Active fees} & \textbf{Total fees} & \textbf{Avg. fee (bps)} \\
    \midrule
    Model~0 & Static & \(35.50\pm 0.30\) & \(54.71\pm 0.19\) & \multicolumn{1}{c}{--} & \(54.71\pm 0.19\) & \(1.00\) (fixed) \\
    Model~0 & Toxicity & \(22.77\pm 0.19\) & \(531.99\pm 4.06\) & \multicolumn{1}{c}{--} & \(531.99\pm 4.06\) & \(14.48\pm 0.08\) \\
    Model~0 & Volatility (DEX) & \(23.85\pm 0.42\) & \(325.14\pm 11.96\) & \multicolumn{1}{c}{--} & \(325.14\pm 11.96\) & \(7.25\pm 0.16\) \\
    Model~0 & Volatility (CEX) & \(25.05\pm 0.46\) & \(247.47\pm 0.59\) & \multicolumn{1}{c}{--} & \(247.47\pm 0.59\) & \(5.33\pm 0.02\) \\
    \midrule
    Model~1 & Static & \(39.11\pm 0.16\) & \(19.55\pm 0.03\) & \(208.99\pm 2.92\) & \(228.53\pm 2.89\) & \(1.00\) (fixed) \\
    Model~1 & Toxicity & \(26.11\pm 0.57\) & \(176.04\pm 8.09\) & \(1348.07\pm 84.13\) & \(1524.11\pm 92.22\) & \(11.72\pm 0.15\) \\
    Model~1 & Volatility (DEX) & \(27.87\pm 0.67\) & \(97.49\pm 0.75\) & \(570.81\pm 9.73\) & \(668.30\pm 10.48\) & \(5.22\pm 0.14\) \\
    Model~1 & Volatility (CEX) & \(27.25\pm 0.61\) & \(94.16\pm 5.23\) & \(950.92\pm 10.97\) & \(1045.09\pm 16.20\) & \(5.36\pm 0.01\) \\
    \midrule
    Model~2 & Static & \(38.53\pm 0.39\) & \(20.63\pm 1.83\) & \(196.85\pm 1.12\) & \(217.47\pm 0.71\) & \(1.00\) (fixed) \\
    Model~2 & Toxicity & \(26.45\pm 1.04\) & \(103.05\pm 4.70\) & \(1200.77\pm 4.53\) & \(1303.82\pm 9.23\) & \(11.01\pm 0.07\) \\
    Model~2 & Volatility (DEX) & \(27.94\pm 0.23\) & \(66.54\pm 4.11\) & \(495.62\pm 1.88\) & \(562.16\pm 2.23\) & \(4.40\pm 0.12\) \\
    Model~2 & Volatility (CEX) & \(26.53\pm 0.27\) & \(82.41\pm 0.25\) & \(854.24\pm 11.71\) & \(936.65\pm 11.97\) & \(5.31\pm 0.02\) \\
    \bottomrule
  \end{tabular}%
  }
\end{table}

\begin{table}[!htbp]
    \centering
    \caption{Arbitrage-related metrics reported as (mean, standard deviation) across simulation runs for the baseline parameterization in Table~\ref{tab:common_params}. “Arb Exec” denotes the number of executed arbitrage trades, “No Arb” the number of instances in which arbitrage was infeasible because prices lay within the no-arbitrage band, and “Arb Rej” the number of arbitrage attempts rejected by the profitability filter.}
    \label{tab:arb_detailed}
    \resizebox{\textwidth}{!}{%
        \begin{tabular}{l cccc cccc cccc}
            \toprule
            & \multicolumn{4}{c}{\textbf{Arb Exec}} & \multicolumn{4}{c}{\textbf{No Arb}} & \multicolumn{4}{c}{\textbf{Arb Rej}} \\
            \cmidrule(lr){2-5} \cmidrule(lr){6-9} \cmidrule(lr){10-13}
            \textbf{Model} & static & vol dex & vol cex & toxicity & static & vol dex & vol cex & toxicity & static & vol dex & vol cex & toxicity \\
            \midrule
            Model 0 & 9020, 75 & 5892, 60 & 6360, 86 & 5843, 104 & 3980, 75& 7108, 59& 6640, 86& 7157, 104& 0.10, 0.30& 0.04, 0.20& 0.10,0.30& 0.06, 0.24\\
            Model 1 & 8696, 62& 6008, 65& 6360, 86& 5529, 97& 4304, 62& 6992, 65& 6640, 86& 7471, 97& 0.16, 0.36& 0.06, 0.23& 0.10, 0.30& 0.12, 0.32\\
            Model 2 & 8680, 62& 6025, 61& 6092, 74& 5518, 99& 4319, 62& 6975, 61& 6908, 73& 7482, 99& 0.08, 0.27& 0.06, 0.23& 0.04, 0.19& 0.10, 0.36\\
            \bottomrule
        \end{tabular}%
    }
\end{table}

\FloatBarrier
\subsection{LVR vs fees vs block size}
In this section we discuss the behaviour of the per-block ratio

\begin{equation}
    R_t = \frac{\Delta \text{LVR}_t}{\Delta F_t} \qquad \text{where} \qquad \Delta \text{LVR}_t = \text{LVR}_t-\text{LVR}_{t-1} ,\quad \Delta F_t=F_t-F_{t-1}
\end{equation}
as the size of the block increases. It quantifies how much the fee earned by the LP offset the accumulated LVR in each block.
The LVR theory implies that holding other factors fixed, the important scaling is \cite{milionis2023}:
\begin{equation}
  \mathbb{E}[\mathrm{LVR} \text{ over an interval of length } \Delta t]
  \;\propto\;
  \sigma^2 \,\Delta t.
  \label{eq:lvr_time_scaling}
\end{equation}
In the simulator, increasing $B$ makes each block represent a longer interval before the end-of-block price
$m_t$ is realized and before the pool is compared against it. This block-size experiment is analogous to latency and market-speed counterfactuals in artificial-market studies of high-frequency trading \cite{leal2016,gao2024}. Larger blocks therefore produce larger typical price
changes within a block (variance scales with time), which increases the magnitude of mispricing and arbitrage 
extraction in that block, consistent with \eqref{eq:lvr_time_scaling}. This naturally produces approximately linear 
growth of $\Delta\text{LVR}_t$ with $B$. However, in general, even if $\Delta\mathrm{LVR}_t$ grows with $B$, the 
ratio $R_t$ will only remain stable if fee revenue also grows proportionally.
Agents' arrivals are parameterized as per-micro-step rates (for both liquidity takers and LP review/refresh events), 
so that the expected number of fee-generating trades and LP actions in a block scales approximately linearly with $B$. 

\begin{figure}[!htbp]
  \centering
  \includegraphics[width=\linewidth]{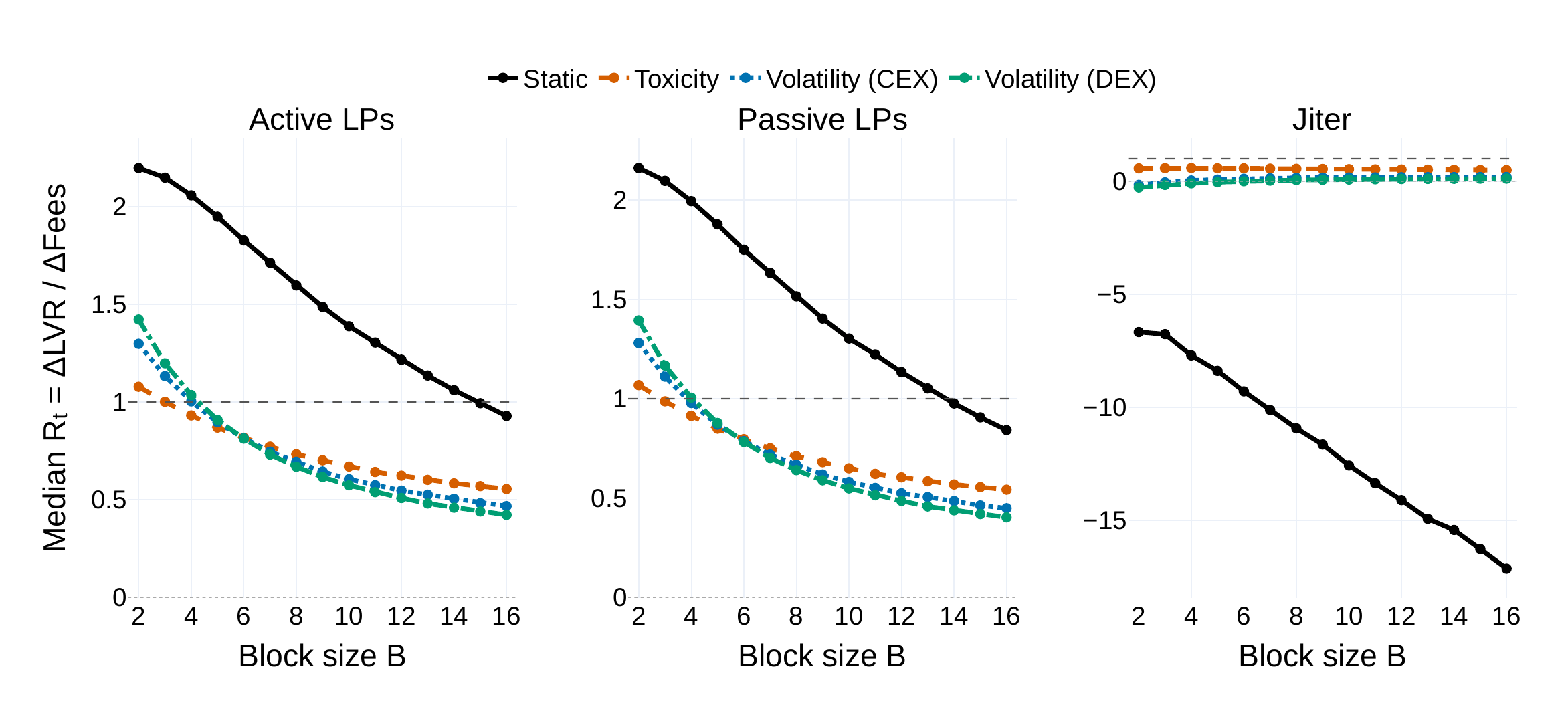}
  \caption{Median of $R_t=\Delta\mathrm{LVR}_t/\Delta F_t$ vs block time $B$ in Model~2 across all four fee schedules 
  (50 runs). Each panel corresponds to one cohort and each line to one fee schedule. The dashed horizontal reference 
  marks $R_t=1$, where marginal fee revenue equals marginal LVR; the dotted reference marks zero. Panels use 
  cohort-specific vertical scales to keep the Jiter series readable. The Jiter curve is computed conditional on 
  successful JIT executions.}
  \label{fig:model2_blocksize_ratio}
\end{figure}

\begin{figure}[!htbp]
  \centering
  \includegraphics[width=\linewidth]{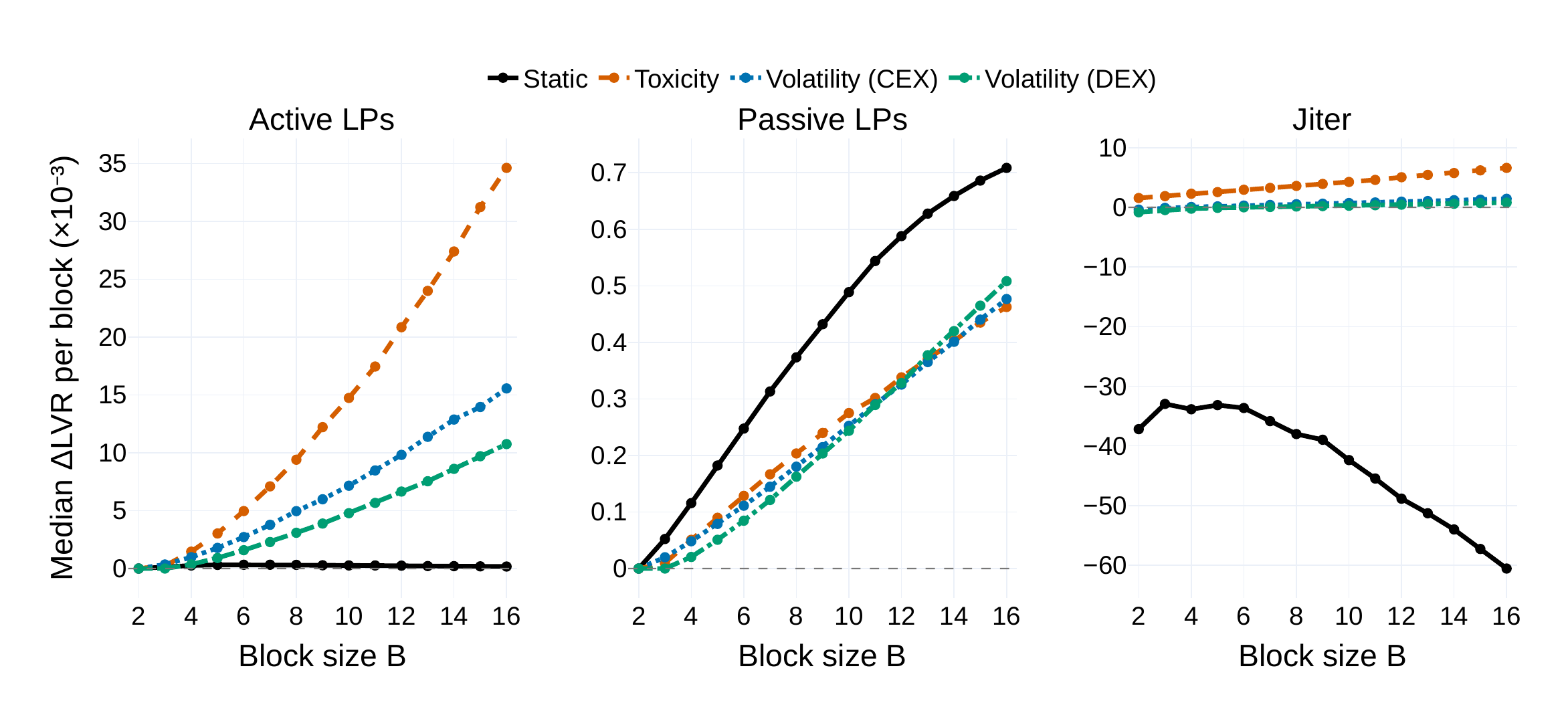}
  \caption{Median per-block $\Delta \mathrm{LVR}$ vs block time $B$ in Model~2 across all four fee schedules (50 runs). 
  Each panel corresponds to one cohort and each line to one fee schedule. The vertical axis reports 
  $10^3\times\Delta \mathrm{LVR}$ to make small per-block changes readable; panels use cohort-specific 
  vertical scales. The dashed horizontal reference marks zero. The Jiter curve is computed conditional on 
  successful JIT executions.}
  \label{fig:delta_lvr_vsd_blocksize}
\end{figure}

Figure~\ref{fig:model2_blocksize_ratio} reports the results of this analysis for the four Model~2 fee schedules. Because the
plotted object is the median of block-level ratios, values below one should be read as improved marginal fee coverage,
not as a guarantee that cumulative hedged PnL over the entire run is positive. The main findings are:
\begin{itemize}
  \item \textbf{Static fee mode:} both passive and active LP cohorts start with median ratios around $2.16$--$2.20$ at 
  $B=2$ and remain above one for almost the entire sweep, crossing below one only at the extreme right tail 
  ($B\approx 14$ for passive LPs and $B\approx 15$ for active LPs). Fees therefore lag LVR over most of the block-size 
  range, and the active cohort is slightly more exposed than the passive one. However, the $\Delta LVR$ per block grows
  for both passive and active LPs, even if for the former the growth is monotone, while for the latter it is positive 
  between $B>2$ and $B<6$ and then starts to decline. This suggests that, as we expected, LVR is mitigated also by just
  having more uniformed flow.
  \item \textbf{Toxicity fee mode:} this is the earliest and cleanest improvement for standing liquidity. The passive 
  and active cohorts are already near break-even at $B=2$ ($1.07$ and $1.08$, respectively), fall below one by 
  $B=3$--$4$, and decline further to about $0.54$--$0.55$ at $B=16$. The toxicity signal therefore raises fees in 
  the same blocks in which stale-price risk is concentrated, rather than only after volatility has accumulated. However,
  $\Delta LVR$ in this case grows monotonically with $B$ for both passive and active LPs, a symptom of the fact that even if 
  the schedule widens the no arbitrage bands, the AMM quotes remain outdated for longer leading to a bigger correction
  when arbitrage occurs. Figure~\ref{fig:delta_lvr_vsd_blocksize} reports the corresponding
  per-block $\Delta LVR$ medians across the fee schedules.
  \item \textbf{Volatility fee modes:} both volatility-based schedules produce a monotone decline in the coverage ratio, 
  but they react more slowly than toxicity. The CEX-volatility rule crosses below one at $B\approx 4$ for passive LPs 
  and $B\approx 5$ for active LPs, while the DEX-volatility rule crosses at about $B\approx 5$ for both cohorts. By 
  $B=16$ the DEX-volatility controller ends slightly lower ($\approx 0.40$--$0.42$) than the CEX-volatility one 
  ($\approx 0.45$--$0.47$), but both remain less targeted than toxicity at short block times.
  \item \textbf{Jiter:} ratios are reported conditional on successful JIT executions. Under static fees the number of 
  successful JIT opportunities is very small, so the estimated ratio is extremely noisy and the median is strongly 
  negative. Under dynamic fees the Jiter ratio is much more stable and generally stays below one: around $0.57 \to 0.49$ 
  under toxicity as $B$ moves from $2$ to $16$, and close to zero at short blocks under the two volatility rules before 
  turning modestly positive at longer blocks. This is consistent with a strategy that captures elevated fees while 
  keeping inventory exposure very short-lived.
\end{itemize}

\FloatBarrier
\section{Conclusions}

This paper introduced a granular agent-based model for studying the economics of Uniswap v3 liquidity provision under 
realistic blockchain frictions and heterogeneous strategic behavior. The model couples a concentrated-liquidity AMM to 
a reference CEX whose mid-price follows a stochastic-volatility Heston dynamics with permanent impact induced by trades, 
and embeds a block-based settlement layer with mempool latency and stochastic transaction ordering. Within this 
environment, arbitrageurs enforce a no-arbitrage band shaped by taker fees and flash-loan costs, while liquidity takers 
split across an uninformed noise-trader component and a best-execution smart router that endogenizes the DEX share as 
fees and liquidity conditions change. Liquidity providers are modeled as cash-budgeted agents who mint/burn positions 
with heterogeneous review clocks, range-selection rules, and risk-management triggers. The framework implements a 
discrete-time rebalancing benchmark that preserves self-financing and inventory matching, enabling a clean decomposition 
of outcomes into fees versus Loss-Versus-Rebalancing (LVR) and thus a direct measurement of “hedged PnL” as the 
economically relevant profitability metric for liquidity provision. Methodologically, the paper contributes to the artificial financial-market literature by treating AMM fee adaptation as a market-design intervention whose effects are mediated by heterogeneous liquidity takers, arbitrageurs, routers, and liquidity providers \cite{raberto2001,chakraborti2011,westerhoff2006,mizuta2020}.

Across the baseline configurations, the simulations confirm a central microstructure mechanism: in a block-based system,
transient within-block deviations between DEX and CEX prices generate systematic arbitrage opportunities that translate
into LVR, especially when liquidity is concentrated near the current price and the AMM effectively posts stale quotes.
In Model 0, a static fee produces persistently negative hedged PnL for passive LPs, despite a non-trivial fraction of
flow routed to the DEX by the smart router. Allowing the taker fee to adapt according to an underlying signal changes
the balance between fee income and adverse selection. Among the tested controllers, the toxicity-based schedule,
constructed from an excess-basis proxy measuring how much the DEX price gap exceeds the contemporaneous fee band,
raises fees specifically in regimes associated with stale-price risk. This generates positive hedged PnL for passive LPs
while maintaining a stable (slightly reduced) DEX market share. The aggregate decomposition indicates that this improvement
comes primarily from higher fee revenue rather than from a lower cumulative LVR: in Model~0, for example, implied LVR is
similar across fee modes and is slightly higher under toxicity than under static fees. Volatility-based schedules also
improve LP outcomes relative to the static benchmark, but generally deliver weaker gains, consistent with volatility being
a noisier and less targeted proxy for the realized adverse-selection channel in this microstructure.

Model 1 highlights a core trade-off created by concentrated liquidity. Introducing active LPs increases depth near 
the prevailing price, improving DEX competitiveness and raising the router’s DEX share; however, the same concentration 
amplifies exposure to LVR, pushing active LP hedged PnL below that of passive LPs under most fee regimes. Notably, 
the toxicity-driven controller is the only specification in which both passive and active LP cohorts achieve positive 
hedged profits, underscoring that where and when fees respond matters more than simply charging higher fees on average. 
In other words, adaptive pricing can be beneficial, but only if the control signal is aligned with the microstructural 
source of adverse selection rather than with broad market variability alone.

Model 2 introduces an MEV searcher performing Just-In-Time (JIT) liquidity. The results emphasize a second,
protocol-design-relevant tension: mechanisms that protect LPs from arbitrage (dynamic fees) can simultaneously increase
the returns to latency/priority strategies that appropriate fee flow. In the static-fee setting, the JIT agent is
approximately break-even, while other LP cohorts remain negative on a hedged basis. Under dynamic fees, the JIT strategy
becomes profitable, because it can time liquidity to capture elevated fees while keeping inventory exposure short-lived.
However, the effect on standing LPs is not uniform. In Model~2, the toxicity controller still makes passive LP
profitability slightly positive (\(18.3\pm 2.8\)), but it no longer suffices to bring active LPs above break-even
(\(-13.6\pm 8.0\)), leaving them approximately break-even or mildly negative. The introduction of JIT liquidity therefore
materially weakens, but does not completely eliminate, the benefit of toxicity-based fees for standing liquidity. This
outcome is economically coherent: when large swaps are systematically wrapped by JIT liquidity, part of the fee stream
that would otherwise compensate longer-horizon LPs is diverted to the searcher precisely during the most lucrative blocks,
leaving residual LPs with weaker protection against LVR.

Overall, the experiments support three broad conclusions. First, LVR is fundamentally a microstructure cost, magnified
by discrete-time execution and latency, and it is better evaluated through modeling the execution layer and not just
via continuous-time diffusion intuition. Second, dynamic fees can improve the sustainability of liquidity provision by
compensating LPs for adverse-selection losses through higher fee income in stale-price-risk states; the present aggregate
results do not imply that positive hedged PnL necessarily comes from reducing cumulative LVR itself. Third, MEV/JIT
dynamics can dominate the distribution of fee revenue, potentially weakening the benefits of adaptive fees for ordinary
LPs unless fee mechanisms and execution rules are co-designed with MEV-resilience in mind.

These findings suggest several directions for future work within the same ABM framework: (i) calibrating agent 
intensities and parameter regimes to specific pools and chains to connect qualitative insights to empirical magnitudes; 
(ii) incorporating explicit gas costs, priority fees, and auction-style ordering to model the full MEV supply chain; 
(iii) extending fee controllers to include anti-JIT features (e.g., time-weighted fee attribution, liquidity “stickiness” 
requirements, or penalties for ultra-short-lived liquidity) while quantifying welfare impacts on takers and routers; 
and (iv) studying multi-pool routing and cross-pool arbitrage, where dynamic fees in one venue propagate strategically 
to others. Taken together, the ABM developed here provides a controlled, interpretable testbed showing that adaptive
pricing is a promising, but not standalone, tool for compensating adverse-selection costs and improving LP economics in
concentrated-liquidity AMMs operating under realistic blockchain microstructure.

\section*{Acknowledgements}
    FL acknowledges support from the grant PRIN2022 DD N. 104 of February 2, 2022 ``Liquidity and systemic risks in centralized and decentralized markets'', codice proposta 20227TCX5W - CUP J53D23004130006 funded by the European Union NextGenerationEU through the Piano Nazionale di Ripresa e Resilienza (PNRR).

\FloatBarrier
\bibliographystyle{abbrv}
\bibliography{bibliography}

\end{document}